\documentclass{iopjournal}

\RequirePackage{graphicx}
\RequirePackage{amsmath,amssymb,amsfonts}
\RequirePackage[numbers,sort&compress]{natbib}
\RequirePackage{hyperref}
\hypersetup{colorlinks=true,linkcolor=blue,citecolor=blue,urlcolor=blue}

\usepackage{keyval,amsfonts, slashed, bm}
\usepackage{textcomp}
\usepackage{ragged2e}
\justifying
\usepackage{MnSymbol}
\usepackage{color}

\definecolor{blue(munsell)}{rgb}{0.2, 0.3, 0.69}
\definecolor{coquelicot}{rgb}{1.0, 0.22, 0.0}
\definecolor{sinopia}{rgb}{0.8,0.25,0.04}
\definecolor{greenopia}{rgb}{0.3,0.65,0.14}

\newcommand{\ba}{\begin{eqnarray}}
\newcommand{\ea}{\end{eqnarray}}

\newcommand{\nn}{\nonumber\\}

\definecolor{sinopia}{rgb}{0.8,0.25,0.04}

\definecolor{greenopia}{rgb}{0.3,0.65,0.14}

\makeatletter
\newsavebox{\@brx}

\begin{document}

\title{Anisotropy effects on heavy quark dynamics in Gribov modified gluon plasma}

\renewcommand{\thefootnote}{\fnsymbol{footnote}}

\author{Sumit $^{1, *}$ 
, 
Jai Prakash$^2$ 
,
Santosh Kumar Das$^{3}$ 
and Najmul Haque $^4$ 
}

\affil{$^1$School of Physics, Beijing Institute of Technology, Beijing 102488, China} \\
\affil{$^2$ School of Physics and Astronomy, Shanghai Key Laboratory for Particle Physics and Cosmology, and Key Laboratory for Particle Astrophysics and Cosmology (MOE), Shanghai Jiao Tong University, Shanghai 200240, China} \\
\affil{$^3$ School of Physical Sciences, Indian Institute of Technology Goa, Ponda-403401, Goa, India} \\
\affil{$^4$School of Physical Sciences, National Institute of Science Education and Research, An OCC of Homi Bhabha National Institute, Jatni-752050, India}


\footnotetext{$^{*}$Author to whom any correspondence should be addressed.}

\email{sumit@ph.iitr.ac.in$^1$, jaiprakashaggrawal2@gmail.com$^2$, santosh@iitgoa.ac.in$^3$, and nhaque@niser.ac.in$^4$}

\keywords{Heavy quarks, Momentum anisotropy, diffusion coefficients, quark-gluon plasma}

\begin{abstract}
\justifying

In the early stages of relativistic heavy-ion collisions, the quark-gluon plasma's momentum distribution is anisotropic, leading to instabilities driven by chromomagnetic plasma modes. In this work, we consider the anisotropic momentum distribution of the medium constituents to investigate its effects on charm and bottom quark dynamics using the nonperturbative Gribov resummation approach within the Fokker-Planck equation framework. Specifically, we investigate the impact of nonperturbative effects and weak anisotropies on the heavy-quark transport coefficients, accounting for the angular dependence of the anisotropy vector relative to the heavy-quark motion direction. Furthermore, the calculated drag and diffusion coefficients are used to estimate heavy-quark energy loss and the nuclear modification factor, accounting for both elastic and inelastic collisions. Our findings indicate that momentum anisotropy, angular dependence, and nonperturbative effects—captured through the scattering amplitudes—play a significant role in determining the transport properties of heavy quarks.

\end{abstract}


\section{INTRODUCTION}\label{section_1}
Relativistic heavy-ion collisions (HIC) produce a transient state of matter known as the quark-gluon plasma (QGP), which remains in local thermodynamic equilibrium for most of its lifetime~\cite{Shuryak:2004cy}. However, the QGP requires a short but finite time to reach this equilibrium state~\cite{Bozek:2010aj}, during which the momentum distribution of the pre-equilibrium plasma is generally anisotropic. The partons that are produced in the early stage of these collisions have momenta along the beam direction, which implies that the longitudinal momenta are enlarged compared to the transverse momenta. Thus, the momentum distribution takes a prolate shape as it becomes strongly elongated along the beam direction. With time, the distribution evolves, and as discussed in Ref.~\cite{Jas:2007rw}, it becomes squeezed along the beam direction, giving an oblate shape. Thus, the characteristic transverse momentum becomes larger than the longitudinal momentum. The QGP state then approaches local equilibrium, although full equilibration is hindered by viscous effects~\cite{Florkowski:2013lya}. Hydrodynamic models, which successfully reproduce many experimental observables in HICs, indicate that the QGP can enter a hydrodynamic regime on a short time scale. However, this should not necessarily be interpreted as complete local thermal equilibrium at very early times, since hydrodynamic behavior may emerge before full thermalization, a process commonly referred to as hydrodynamization~\cite{Bozek:2010aj}. Collective excitations in the early phases of QGP are central to understanding the thermalization scenario; see Refs.~\cite{Kurkela:2011ti, Kurkela:2011ub, Attems:2012js} for early studies and Refs.~\cite{Ipp:2010uy, Guin:2025lpy, Mrowczynski:2016etf} for further discussions. \\
Heavy flavor particles~\cite{Rapp:2018qla, Das:2024vac, Das:2024wht} are among the important probes used to study the transport properties of $\mathrm{QGP}$ because they are produced at very early times in relativistic $\mathrm{HIC}$. Thus, the heavy quarks $\mathrm{(HQs)}$ interact with the bulk medium from the very early stages up to the final stages of the evolution, thereby carrying valuable information about these interactions. This interaction information is generally encoded in experimental observables such as the nuclear modification factor ($\mathrm{R_{AA}}$) and the elliptic flow ($\mathrm{v}_{2}$). Experimentally, it has been observed that both $\mathrm{R_{AA}}$ and $\mathrm{v}_{2}$ at large $\mathrm{p_{T}}$  of heavy flavor particles are similar to those of light flavor particles, which can be partly understood within perturbative QCD frameworks that include radiative and collisional energy loss. However, the sizable $\mathrm{v}_{2}$ of $\mathrm{HQs}$, particularly at low and intermediate $\mathrm{p_{T}}$, strongly suggests that $\mathrm{HQs}$ interact substantially with the medium and participate in its collective flow~\cite{CMS:2017qjw, ALICE:2021rxa, STAR:2018zdy}. Due to their significantly larger masses compared to light quarks, heavy-flavor particles are expected to thermalize more slowly in hot QCD matter. Thus, hydrodynamics, which are very successful at describing the dynamics of bulk matter, do not apply to the study of heavy particles. Within the assumption of soft momentum transfer, the relativistic non-linear Boltzmann equation converts to the Fokker-Planck equation using the diffusion approximation, and also within this approximation, a heavy flavor particle can be treated as a Brownian particle~\cite{Svetitsky:1987gq}.\\
To describe the time evolution of heavy flavor particles in hot QCD matter, Langevin or Boltzmann transport approaches are utilized under the assumption that the colliding partners from the $\mathrm{QGP}$ are in thermal equilibrium~\cite{Svetitsky:1987gq, Moore:2004tg, vanHees:2004gq, vanHees:2005wb}. Several recent implementations are described in Refs.~\cite{Zhao:2020jqu, He:2014cla, Das:2015ana, Song:2015sfa, Cao:2018ews}. However, as discussed, this assumption is not always true in hot QCD matter for the models to which it is applied, for several reasons. (a) Bulk particles also require a finite time to thermalize after their production is complete. Also, color-glass condensate~\cite{Iancu:2000hn} theory predicts that the transverse pressure is greater than the longitudinal pressure in the early stages of the $\mathrm{HIC}$. (b) As a consequence of hard scattering at production, the transverse momentum spectra of particles fall off like a power law in $\mathrm{HIC}$ at large momenta~\cite{Tsallis:1987eu, Song:2012at}. (c) Even during the central collisions, there is a corona of nucleons that do not sufficiently scatter to reach the thermal equilibrium and thus can modify the transport coefficients as shown in Ref.~\cite{Aichelin:2010ns}. Therefore, in order to compare the calculated transport coefficients with experimental data, it is necessary to include these nonequilibrium effects. Thus, it is helpful to estimate the transport coefficients of $\mathrm{HQs}$ when the $\mathrm{QGP}$ is not in complete thermal equilibrium. $\mathrm{HQ}$ transport properties in equilibrium QCD media have been extensively studied within perturbative energy-loss frameworks~\cite{Braaten:1991jj,Wang:1994fx, Gyulassy:1993hr, Klein:1998du,Braaten:1991we}. Further developments, including improved treatments of collisional and radiative transport, can be found in Refs.~\cite{Mustafa:1997pm, Dokshitzer:2001zm, Mustafa:2004dr, Qin:2007rn, Mazumder:2011nj} (Also, see for recent updates~\cite{Abir:2012pu, Abir:2011jb, Ruggieri:2022kxv, Das:2022lqh, Prakash:2023hfj}). Anisotropic QCD media have also been investigated in the context of $\mathrm{HQ}$ transport~\cite{Das:2012ck, Shaikh:2021lka, Pandey:2025rxv, Kumar:2021goi, Prakash:2023wbs}, energy-loss phenomena~\cite{Carrington:2015xca, Song:2019cqz, Pandey:2023dzz, Boguslavski:2023fdm, Boguslavski:2020tqz} along with other nonequilibrium scenarios~\cite{Carrington:2014bla, Singh:2019cwi, Thakur:2021vbo, Thakur:2020ifi} and  (see also~\cite{Du:2024riq, Guo:2024mgh, Dong:2022mbo, Srivastava:2015via}). Additional studies on anisotropic transport dynamics and collective modes can be found in Refs.~\cite{Romatschke:2003ms, Kumar:2017bja, Jamal:2017dqs, Ghosh:2020sng, Zhao:2023mrz}. Recently, the $\mathrm{HQs}$ diffusion coefficients have been studied in the pre-equilibrium Glasma phase~\cite{Mrowczynski:2017kso, Carrington:2022bnv, Ruggieri:2018rzi, Das:2017dsh, Khowal:2021zoo} (See also \cite{Boguslavski:2020tqz}).\\
As discussed earlier, there are strong correlations that have been observed experimentally~\cite{CMS:2017qjw, ALICE:2021rxa, STAR:2018zdy} among the heavy flavor particles and the medium constituents, which show that one needs to go beyond perturbative calculations~\cite{Moore:2004tg,vanHees:2004gq,vanHees:2005wb} to interpret the experimental data. Several models take into account the nonperturbative aspects~\cite{vanHees:2007me} through the quasi-hadronic bound states with sequential hadronization from coalescence and fragmentation~\cite{Greco:2003xt, Zhao:2023nrz, Elfner:2022iae}. Dynamical quasi-particle models are also helpful in studying the transport properties, which take into account the nonperturbative contribution through the lattice thermodynamics results~\cite{Gossiaux:2008jv, Plumari:2011mk, Scardina:2017ipo, Berrehrah:2014kba}. Another strategy for including nonperturbative effects in $\mathrm{HQ}$ dynamics can be achieved via the Gribov-Zwanziger action~\cite{Gribov:1977wm, Zwanziger:1989mf}. This description includes a scale $g^{2}T$ known as a (chromo)magnetic scale to include the nonperturbative effects since the perturbative expansion of finite temperature Yang-Mills theory breaks down at this scale, known as the Linde problem~\cite{Linde:1980ts}. This nonperturbative resummation approach fixes the residual gauge discrepancy in the infrared sector of QCD that persists after Faddeev-Popov quantization, famously known as the Gribov problem~\cite{Vandersickel:2012tz}. At zero temperature, this approach, with the consideration of the dynamical mass term in gluon and ghost propagator, is quite successful in reproducing the lattice results~\cite{Tissier:2010ts, Mintz:2018hhx}. For more details on this, look at the recent review~\cite{Pelaez:2021tpq}. \\
Gribov's quantization approach has been generalized at finite temperature by Zwanziger, and it has been shown that the Gribov mass parameter $\gamma_{G}$ behaves as a magnetic scale in the asymptotically high-temperature limit~\cite{Zwanziger:2004np, Zwanziger:2006sc}. Recently, several studies have explored the properties of the deconfined state of matter within the Gribov framework and its implications for QCD observables. Early investigations of thermodynamic and transport properties can be found in Refs.~\cite{Fukushima:2013xsa, Su:2014rma, Kharzeev:2015xsa, Florkowski:2015dmm, Florkowski:2015rua}. Subsequent developments addressing transport phenomena and related observables are discussed in Refs.~\cite{Jaiswal:2020qmj, Bandyopadhyay:2015wua, Sumit:2023hjj, Bandyopadhyay:2023yjp, Madni:2024xyj}, while more recent applications can be found in Refs.~\cite{Madni:2024ubw, Du:2024sbv}. For the $\mathrm{HQ}$ phenomenlogy which we are interested in here, $\mathrm{HQ}$ potential has been calculated using Gribov approach~\cite{Debnath:2023dhs,Wu:2022nbv}, $\mathrm{HQ}$ diffusion coefficient using langevin and Fokker-Planck approach has been studied in Ref.~\cite{Madni:2022bea,Sumit:2023oib}, extending the $\mathrm{HQ}$ transport coefficients from eikonal to non eikonal scenario in gluon radiation~\cite{Mazumder:2024grc} and the energy loss of fast moving parton has been studied in Ref.~\cite{Debnath:2023zet}.\\
In this work, we have extended the $\mathrm{HQ}$ momentum evolution, from isotropic~\cite{Sumit:2023oib} to anisotropic $\mathrm{QCD}$ medium and studied the outcomes of momentum anisotropy on $\mathrm{HQ}$ transport coefficients, energy loss, and the $\mathrm{R_{AA}}$ of $\mathrm{HQs}$, using the Gribov-Zwanziger action approach, thus including the nonperturbative aspects in the $\mathrm{HQ}$ dynamics. In the anisotropic $\mathrm{QCD}$ medium, using the tensor basis, the decomposition of $\mathrm{HQ}$ drag coefficient and diffusion coefficient leads to two drag and four diffusion coefficients of $\mathrm{HQs}$, contrary to the one drag and two diffusion coefficients in the isotropic medium. The interaction of the $\mathrm{HQs}$ with the medium constituents is encoded in the matrix elements of the collisional and radiative processes. Gribov propagator provides a natural infrared cut-off in the matrix elements calculation, compared to perturbative approaches, which handle the $\mathrm{IR}$ divergences present in $t$-channel exchange diagrams. Momentum anisotropic effects are included through the nonequilibrium part of the distribution of the constituent particles of the hot $\mathrm{QCD}$ medium. We have quantified the effects of momentum anisotropy through the temperature and momentum dependence of $\mathrm{HQ}$ transport coefficients by combining both the collisional and radiative processes. It has been deduced that the arbitrary orientation of the direction of anisotropy with respect to $\mathrm{HQ}$ momentum and the anisotropy strength plays an important role in quantifying the drag and diffusion properties of $\mathrm{HQs}$. \\ 
The work is sketched as follows. After this brief introduction in section~\ref{section_1}, we discuss the formalism of the $\mathrm{HQ}$ drag and diffusion coefficients in the isotropic and anisotropic hot $\mathrm{QCD}$ matter in section~\ref{section_2}. We present results for the $\mathrm{HQ}$ drag and diffusion coefficients, utilizing the scattering elements calculated using the Gribov propagator and energy loss of $\mathrm{HQs}$, combining both the collisional and radiative contributions along with the experimental observable $\mathrm{R_{AA}}$ in section~\ref{section_3}. Finally, we summarize the work in section~\ref{section_4}. We quote the scattering elements for elastic processes in the appendix~\ref{appendix_A}, and different projections between the anisotropic vector and the $\mathrm{HQ}$ momentum in the center of mass frame $\mathrm{(COM)}$ are presented in appendix~\ref{appendix_B} for completeness.
\section{FORMALISM: HEAVY QUARK DRAG AND DIFFUSION IN HOT QCD MEDIUM}
\label{section_2}
The motion of the $\mathrm{HQs}$ embedded in a thermal medium can be described through the well-known Fokker-Planck equation described as~\cite{Svetitsky:1987gq} 
\begin{equation}\label{FP_equa.}
\frac{\partial \mathfrak{f}_{\mathrm{HQ}}}{\partial t}=\frac{\partial}{\partial p_i}\left[\mathcal{A}_i(\bm{p}) \mathfrak{f}_{\mathrm{HQ}}+\frac{\partial}{\partial p_j}\left[\mathcal{B}_{i j}(\bm{p}) \mathfrak{f}_{\mathrm{HQ}}\right]\right] \,.
\end{equation}
Here, $\mathfrak{f}_{\mathrm{HQ}}$ represents the momentum distribution of $\mathrm{HQs}$ in the thermal medium and $\mathcal{A}_{i}(\bm{p})$, $\mathcal{B}_{i j}(\bm{p})$ denote the drag and diffusion coefficients of the $\mathrm{HQ}$ respectively.  We can derive the Fokker-Planck equation from the master Boltzmann equation by applying the Landau approximation. This approximation assumes that the scatterings between the partons in the medium can be considered soft, i.e., the rate of collisions which changes the momentum of the  $\mathrm{HQ}$ from $\bm{p}$ to $(\bm{p}-\bm{k})$ falls off rapidly with the $\left|\bm{k}\right|$. The validity of the Fokker-Planck/Langevin description, therefore, relies on the dominance of multiple soft scatterings and remains within the usual diffusion approximation. In the following section, we discuss collisional processes in isotropic and anisotropic media, and later consider radiative processes in separate subsections for convenience. 
\subsection{Collisional Processes: Isotropic QCD medium}
Let us consider the two body elastic interaction processes, i.e., $\mathrm{HQ}(P)+ l(Q) \rightarrow$ $\mathrm{HQ}\left(P^{\prime}\right)+l\left(Q^{\prime}\right)$,  where $l$ corresponds to the light particles, i.e. light quarks, light antiquarks, and gluons. The drag and diffusion tensor can be written in the following manner
\begin{eqnarray}
\mathcal{A}_i&= & \frac{1}{2 E_p} \int \frac{d^3 \bm{q}}{(2 \pi)^3 2 E_q} \int \frac{d^3 \bm{q}^{\prime}}{(2 \pi)^3 2 E_{q^{\prime}}} \int \frac{d^3 \bm{p}^{\prime}}{(2 \pi)^3 2 E_{p^{\prime}}} \frac{1}{g_{\mathrm{HQ}}} (2 \pi)^4 \delta^4\left(P+Q-P^{\prime}-Q^{\prime}\right) \nonumber\\
&\times & \sum\left|\mathcal{M}_{2 \rightarrow 2}\right|^2 \mathfrak{f}_k({E_{q}})\left[1+a_k \mathfrak{f}_k\left({E_{q^{\prime}}}\right)\right]\left[\left(\bm{p}-\bm{p}^{\prime}\right)_i\right]=\llangle\left(\bm{p}-\bm{p}^{\prime}\right)_i\rrangle , \hspace*{-.0cm} 
\label{drag_definition} 
\end{eqnarray}
and,
\begin{equation}
\begin{aligned}
\mathcal{B}_{i j}= & \frac{1}{2}\llangle\left(\bm{p}-\bm{p}^{\prime}\right)_i\left(\bm{p}-\bm{p}^{\prime}\right)_j
\rrangle \,.
\end{aligned}\label{diffusion_definition}
\end{equation} 
Here, $g_{\mathrm{HQ}}$ denotes the $\mathrm{HQ}$ statistical factor and $\mathfrak{f}_k$ denotes the near-equilibrium distribution function. Additionally, the delta function ensures energy conservation in the two-body scattering process. The scattering amplitude $\left|\mathcal{M}_{2 \rightarrow 2}\right|^2$ result is presented in the appendix~\ref{appendix_A}. The term $\left(1+a_k\mathfrak{f}_k\left({E_{q^{\prime}}}\right)\right)$ represents the Fermi suppression and Bose enhancement regarding the final state particles' phase space. From the Eqs.~\eqref{drag_definition} and \eqref{diffusion_definition}, it can be seen that the drag coefficient corresponds to the thermal average of the momentum transfer $\left(\bm{p}-\bm{p}^{\prime}\right)$ as an artifact of medium interactions. On the other hand, diffusion coefficients are defined as the average squared momentum transfer per unit time. Since in the isotropic medium, drag and diffusion tensor depend on the $\mathrm{HQ}$ momentum only, thus, after using tensorial properties, one obtains the drag coefficient $\mathcal{A}$ as    
\begin{equation}
\mathcal{A}=\llangle 1\rrangle-\frac{\llangle\bm{p} \cdot \bm{p}^{\prime}\rrangle}{p^2} \,.
\end{equation}
While transverse $\mathcal{B}_{0}$ and longitudinal $\mathcal{B}_{1}$ diffusion coefficient becomes
\begin{equation}
\mathcal{B}_0=\frac{1}{4}\left[\llangle p^{\prime 2}\rrangle-\frac{\llangle\left(\bm{p}^{\prime} \cdot \bm{p}\right)^2\rrangle}{p^2}\right]\, , \quad \quad \quad
\mathcal{B}_1=\frac{1}{2}\left[\frac{\llangle\left(\bm{p}^{\prime} \cdot \bm{p}\right)^2\rrangle}{p^2} - 2\llangle\bm{p}^{\prime} \cdot \bm{p}\rrangle+p^2\llangle1\rrangle\right] \,.
\end{equation}
Note that the transverse and longitudinal components are identified relative to the $\mathrm{HQ}$ momentum. For simplicity, the kinematics is studied in the $\mathrm{COM}$ frame and the average of a generic function $\mathcal{F}(p)$ in $\mathrm{COM}$ is given by 
\ba
\llangle \mathcal{F}(\bm{p})\rrangle &= & \frac{1}{\left(512 \pi^4\right) E_p g_{\mathrm{HQ}}} \int_0^{\infty} q\, d q  \int_0^\pi d \chi \sin \chi \int_0^\pi d \Theta_{\mathrm{cm}} \sin \Theta_{\mathrm{cm}} \int_0^{2 \pi} d \Phi_\mathrm{cm} \left(\frac{s-M_{\mathrm{HQ}}^2}{s}\right) \nn 
&\times&  \mathfrak{f}_k\left(E_q\right) \left[1+a_k \mathfrak{f}_k\left(E_{q^{\prime}}\right)\right] \sum\left|\mathcal{M}_{2 \rightarrow 2}\right|^2  \mathcal{F}(\bm{p}) \, .
\label{F(p)_def.}
\ea
Here, the angle $\chi$ defines the relative orientation between the $\mathrm{HQ}$ and medium particles in the lab frame, whereas $\Theta_{\mathrm{cm}}$ and $\Phi_{\mathrm{cm}}$ represent the polar and azimuthal angles, respectively, in the $\mathrm{COM}$ frame. The Mandelstam variables $s$, $t$, and $u$ are expressed in terms of particle momenta as:
\ba
s & =& (P+Q)^{2} = \left(E_p+E_q\right)^2-\left(p^2+q^2+2 p q \cos \chi\right) \, ,\quad \quad \quad
t  = (P^{\prime}-P)^{2} = 2 p_{\mathrm{cm}}^2\left(\cos \Theta_{\mathrm{cm}}-1\right) \, , \nn
u & =& (P^{\prime}-Q)^{2} = 2 M_{\mathrm{HQ}}^2-s-t \, .
\ea
Here $p_{\mathrm{cm}} = |\bm{p}_{\mathrm{cm}}| $ is the magnitude of the $\mathrm{HQ}$ momentum before the collision in the $\mathrm{COM}$ frame. In order to determine the drag and diffusion coefficients, one needs to evaluate the quantity $\left(\bm{p}\cdot\bm{p}^{\prime}\right)$. To that end, the Lorentz transformation, which connects the lab and the $\mathrm{COM}$ frame is given by $ \bm{p}^{\prime}=\gamma_{\mathrm{cm}}\left(\bm{p}_{\mathrm{cm}}^{\prime}+\bm{v}_{\mathrm{cm}} E_{\mathrm{cm}}^{\prime}\right),$ where  $\gamma_{\mathrm{cm}}=(E_p+E_q)/\sqrt{s}$ and the velocity in the $\mathrm{COM}$ is given by $\bm{v}_{\mathrm{cm}}=(\bm{p}+\bm{q})/(E_p+E_q)$. Within $\mathrm{COM}$ frame, one can decompose $\bm{p}_{\mathrm{cm}}^{\prime}$ as 
\begin{equation}
\begin{aligned}
\bm{p}_{\mathrm{cm}}^{\prime} & = p_{\mathrm{cm}}\left(\cos \Theta_{\mathrm{cm}} \bm{x}_{\mathrm{cm}}+\sin \Theta_{\mathrm{cm}} \sin \Phi_{\mathrm{cm}} \bm{y}_{\mathrm{cm}} \right. + \left. \sin \Theta_{\mathrm{cm}} \cos \Phi_{\mathrm{cm}} \bm{z}_{\mathrm{cm}}\right),
\end{aligned}
\end{equation}
where $p_{\mathrm{cm}}=(s-M_{\mathrm{HQ}}^2)/(2 \sqrt{s})$ is the $\mathrm{HQ}$ momentum and $E_{\mathrm{cm}}= \left(p_{\mathrm{cm}}^2+M_{\mathrm{HQ}}^2\right)^{1 / 2}$ is the energy of the $\mathrm{HQ}$ in the $\mathrm{COM}$ frame. The axes $\bm{x}_{\mathrm{cm}}, \bm{y}_{\mathrm{cm}}$, and $\bm{z}_{\mathrm{cm}}$ are defined as follows $\bm{x}_{\mathrm{cm}} = \bm{p}_{\mathrm{cm}}/ p_{\mathrm{cm}} , \, \bm{y}_{\mathrm{cm}} = N^{-1} \left(\bm{v}_{\mathrm{cm}} \right.
- \left. (\bm{p}_{\mathrm{cm}}\cdot \bm{v}_{\mathrm{cm}})\bm{p}_{\mathrm{cm}}/p_{\mathrm{cm}}^{2} \right) ,\, \bm{z}_{\mathrm{cm}}  = \bm{x}_{\mathrm{cm}} \times \bm{y}_{\mathrm{cm}}$ where $N^2 = v_{\mathrm{cm}}^{2} - (\bm{p}_{\mathrm{cm}} \cdot \bm{v}_{\mathrm{cm}})^{2}/ p_{\mathrm{cm}}^{2}$. Using these definitions, one can find that
\begin{equation}
\begin{aligned}
\bm{p} \cdot \bm{p}^{\prime} & = E_p E_p^{\prime}-\hat{E}_{\mathrm{cm}}^2+\hat{p}_{\mathrm{cm}}^2 \cos \Theta_{\mathrm{cm}} \, .
\end{aligned}
\end{equation}
\subsection{Collisional Processes: Anisotropic QCD medium}
The momentum distribution of quarks and gluons, which is anisotropic in momentum space, plays a crucial role in determining the state of the medium. An ansatz has been introduced in Ref.~\cite{Romatschke:2003ms} to incorporate momentum anisotropy by modifying the isotropic distribution function through a rescaling—either elongation or contraction, along a preferred direction, may be written in the form
\begin{equation}
\begin{aligned}
\mathfrak{f}_{\xi}(\bm{p})= \, \mathfrak{f}_{0}\left(\sqrt{\bm{p}^2+\xi(\bm{p} \cdot \bm{n})^2}\right),
\label{f_aniso}
\end{aligned}
\end{equation}
where $\mathfrak{f}_{0}(|\bm{p}|)$ is an isotropic distribution of medium constituents, and the parameter, $\xi \in(-1, \infty)$ determines the shape of the distribution. When $-1<\xi<0$, the momentum distribution is elongated in the direction of $\bm{n}$, resulting in a prolate shape. Conversely, for $\xi>0$, the distribution is squeezed along the unit vector $\bm{n}$, becoming increasingly oblate as $\xi$ increases. Note that the number density in the local rest frame can be factorized into a function that depends solely on $\xi$ and another function which depends solely on the scale $T$ as $n_\xi(T)=n_0(T)/\sqrt{1+\xi} $~\cite{Alqahtani:2017mhy}. Since the unnormalized Romatschke-Strickland (RS) distribution mentioned in Eq.~\eqref{f_aniso} does not impose a matching condition, the isotropic and anisotropic systems at the same hard scale generally differ in number, energy, and entropy density. Thus, the present framework should be understood as a fixed-hard-scale setup. The implications of this choice for the numerical results are discussed in Sec.~\ref{section_3}. We consider a weakly anisotropic system here, i.e., $\xi\ll 1$. For the medium constituent of species $k$, we identify $\mathfrak{f}_k^{(a)}({\bm q})\equiv \mathfrak{f}_{\xi}({\bm q})$, which gives $\mathfrak{f}_k^{(a)}=\mathfrak{f}_k^0+\delta \mathfrak{f}_k$ with~\cite{Srivastava:2015via}
\begin{equation}
\begin{aligned}
\delta \mathfrak{f}_k=-\frac{\xi}{2 E_q T}(\bm{q} \cdot \bm{n})^2\left(\mathfrak{f}_k^0\right)^2 \exp \left(E_q/T\right).
\label{aniso_dist._func._devi.}
\end{aligned}
\end{equation}
In Eq.~\eqref{aniso_dist._func._devi.}, the parameter $T$ represents an effective hard-momentum scale characterizing the typical parton momentum, which reduces to the equilibrium temperature in the isotropic limit within the RS framework~\cite{Strickland:2014eua}. Thus, $T$ is used here as a convenient scale parameter rather than implying local thermal equilibrium. Since only linear terms in $\xi$ are retained, the present framework should be interpreted as a weak-anisotropy expansion, applicable to perturbatively small deviations from isotropy. Also note that the resulting angular structure $({\bm q}\cdot{\bm n})^2$ is closely related to shear-viscous corrections in hydrodynamic treatments. However, here $\xi$ is treated as an external parameter rather than dynamically evolved. The superscripts $\text{`a'}$ and $`0$' refer to anisotropic and isotropic media, respectively. Because of the unit vector in an anisotropic medium, contrary to the isotropic one, the drag coefficient can now be decomposed into two different orthogonal vectors as 
\begin{equation}
\begin{aligned}
\mathcal{A}_i=p_i \mathcal{A}_0^{(\text{a)}}+\tilde{n}_i \mathcal{A}_1^{(\text{a})}
\end{aligned}
\end{equation}
where $\tilde{n}^i=\left(\delta_{i j}-\frac{p_i p_j}{p^2}\right) n^j$ such that $\bm{p} \cdot \tilde{\bm{n}}=0$. Thus, the two different components of the drag coefficient become
\begin{equation}
\mathcal{A}_0^{(\text{a})}=p_i \mathcal{A}_i / p^2=\llangle 1\rrangle-\frac{\left\llangle\bm{p} \cdot \bm{p}^{\prime}\right\rrangle}{p^2}, \quad \quad \quad \quad \mathcal{A}_1^{(\text{a})}=\tilde{n}_i \mathcal{A}_i / \tilde{n}^2=-\frac{1}{\tilde{n}^2}\left\llangle\tilde{\bm{n}} \cdot \bm{p}^{\prime}\right\rrangle.
\end{equation}
The new component $\mathcal{A}_1^{\text{(a)}}$ consists only anisotropic contribution to the $\mathrm{HQ}$ drag coefficient. The average of a function $\mathcal{F}\left(p^{\prime}\right)$ in anisotropic medium can be written as 
\begin{equation}
\begin{aligned}
\left\llangle\mathcal{F}\left(p^{\prime}\right)\right\rrangle=\left\llangle \mathcal{F}\left(p^{\prime}\right)\right\rrangle_0+\left\llangle\mathcal{F}\left(p^{\prime}\right)\right\rrangle_{\mathrm{a}},
\label{F(p)_aniso_medium}
\end{aligned}
\end{equation}
where the isotropic contribution $\llangle \mathcal{F}(\bm{p})\rrangle_0$ is defined in Eq.~\eqref{F(p)_def.} and the anisotropic contribution $\llangle \mathcal{F}(\bm{p})\rrangle_\mathrm{a}$  can be written analogous to Eq.~\eqref{F(p)_def.}, using the anisotropic distribution function definition, as    
\begin{align}
\left\llangle \mathcal{F}(\bm{p})\right\rrangle_\mathrm{a}&= \frac{1}{1024 \pi^5\, E_p g_{\mathrm{HQ}}} \int_0^{\infty} d q \,q\left(\frac{s-M_{\mathrm{HQ}}^{2}}{s}\right)  \int_0^\pi d \chi \sin \chi \int_0^{2 \pi} d \Phi \int_0^\pi d \Theta_{c m} \sin \Theta_{c m} \nonumber\\
& \times \int_0^{2 \pi} d \Phi_{c m}\Big\{\delta \mathfrak{f}_k(\bm{q})\left[1+a_k \mathfrak{f}_k^0\left(\bm{q}^{\prime}\right)\right]  +a_k \mathfrak{f}_k^0(\bm{q}) \delta \mathfrak{f}_k\left(\bm{q}^{\prime}\right)\Big\} \mathcal{F}(\bm{p}) \sum\left|\mathcal{M}_{2 \rightarrow 2}\right|^2 .
\end{align}
Similar to Eq.~\eqref{F(p)_aniso_medium}, the component of drag force can be written similarly as
\begin{equation}
\mathcal{A}_0^{(\text{a})}=\mathcal{A}_0+\delta \mathcal{A}_0,
\end{equation}
having $\mathcal{A}_{0}$ as isotropic contribution and $\delta \mathcal{A}_0$ refers to the anisotropic correction to the drag coefficient. Now for the $\mathrm{HQ}$ diffusion tensor $\mathcal{B}_{ij}$, the orthogonal basis can be constructed using the momentum and anisotropy vector as~\cite{Romatschke:2003ms}
\begin{align}
\mathcal{B}_{i j}= & \left(\delta_{i j}-\frac{p_i p_j}{p^2}\right) \mathcal{B}_0^{(\text{a)}}+\frac{p_i p_j}{p^2} \mathcal{B}_1^{(\text{a})}  +\frac{\tilde{n}_i \tilde{n}_j}{\tilde{n}^2} \mathcal{B}_2^{(\text{a)}}+\left(p^i \tilde{n}^j+p^j \tilde{n}^i\right) \mathcal{B}_3^{(\text{a})}.
\label{diffusion_aniso_medium}
\end{align}
The different diffusion components can be obtained via the proper projection of Eq.~\eqref{diffusion_aniso_medium} as~\cite{Kumar:2021goi} 
\begin{subequations}
\begin{align}
\mathcal{B}_0^{(\text {a})} & =\frac{1}{2}\left[\left(\delta_{i j}-\frac{p_i p_j}{p^2}\right)-\frac{\tilde{n}_i \tilde{n}_j}{\tilde{n}^2}\right] \mathcal{B}_{i j} =\frac{1}{4}\left[\llangle p^{\prime 2}\rrangle-\frac{\left\llangle\left(\bm{p}^{\prime} \cdot \bm{p}\right)^2\right\rrangle}{p^2}-\frac{\left\llangle\left(\bm{p}^{\prime} \cdot \tilde{\bm{n}}\right)^2\right\rrangle}{\tilde{n}^2}\right], \\
\mathcal{B}_1^{(\text{a})} & =\frac{p_i p_j}{p^2} \mathcal{B}_{i j} =\frac{1}{2}\left[\frac{\left\llangle\left(\bm{p}^{\prime} \cdot \bm{p}\right)^2\right\rrangle}{p^2}-2\left\llangle\left(\bm{p}^{\prime} \cdot \bm{p}\right)\right\rrangle+p^2\llangle 1\rrangle\right], \\
\mathcal{B}_2^{(\text{a})} & =\left[\frac{2 \tilde{n}_i \tilde{n}_j}{\tilde{n}^2}-\left(\delta_{i j}-\frac{p_i p_j}{p^2}\right)\right] \mathcal{B}_{i j} =\frac{1}{2}\left[-\left\llangle p^{\prime 2}\right\rrangle+\frac{\left\llangle\left(\bm{p}^{\prime} \cdot \bm{p}\right)^2\right\rrangle}{p^2}+\frac{2\left\llangle\left(\bm{p}^{\prime} \cdot \tilde{\bm{n}}\right)^2\right\rrangle}{\tilde{n}^2}\right], \\
\mathcal{B}_3^{(\text{a})} & =\frac{1}{2 p^2 \tilde{n}^2}\left(p^i \tilde{n}^j+p^j \tilde{n}^i\right) \mathcal{B}_{i j} =\frac{1}{2 p^2 \tilde{n}^2}\left[-p^2\left\llangle\left(\bm{p}^{\prime} \cdot \tilde{\bm{n}}\right)\right\rrangle+\left\llangle\left(\bm{p}^{\prime} \cdot \bm{p}\right)\left(\bm{p}^{\prime} \cdot \tilde{\bm{n}}\right)\right\rrangle\right] .
\end{align}
\label{diffusion_components}
\end{subequations}
Here, $\tilde{n}^2=1-((\bm{p} \cdot \hat{\bm{n}})^2 /{p^2})=1-\cos^2 \Theta_n$. To determine the quantity in $\mathrm{COM}$ frame
$\llangle\tilde{\bm{n}} \cdot \bm{p}^{\prime}\rrangle$, one can consider $\bm{n}=\left(\sin \Theta_n, 0, \cos \Theta_n\right)$, where angle $\Theta_n$ is the angle between anisotropy vector and the $\mathrm{HQ}$ momentum direction which is chosen as $\bm{p}=(0,0, p)$. Also, light quark momentum can be decomposed as $\bm{q}=(q \sin \chi \cos \Phi, 
q \sin \chi \sin \Phi, q \cos \chi)$. Thus we have
\begin{equation}
\begin{aligned}
& \bm{p} \cdot\bm{q}=p q \cos \chi \, , \quad \quad \quad \bm{p} \cdot \bm{n}=p \cos \Theta_n, \quad \quad \quad \bm{q} \cdot \bm{n}=q \sin \chi \cos \Phi \sin \Theta_n+q \cos \chi \cos \Theta_n.
\end{aligned}
\end{equation} 
Further, using the definition of $\tilde{n}^i$, we can define
\begin{equation}
\begin{aligned}
\left\llangle\tilde{\bm{n}} \cdot \bm{p}^{\prime}\right\rrangle=\left\llangle\bm{n} \cdot \bm{p}^{\prime}\right\rrangle-\left\llangle\bm{p} \cdot \bm{p}^{\prime}\right\rrangle \frac{\cos \Theta_n}{p} .
\label{n_tilde_p_prime}
\end{aligned}
\end{equation}
After doing some simplification, the Eq.~\eqref{n_tilde_p_prime} takes the form
\begin{equation}
\begin{aligned}
\tilde{\bm{n}} &\cdot \bm{p}^{\prime}  =  \frac{\gamma_{c m}}{1+\gamma_{c m}^2 v_{c m}^2}\bigg\{{p}_{c m}\left[\cos \Theta_{c m}\left({\bm{x}}_{cm} \cdot\bm{n}\right)+\sin \Theta_{c m} \sin \Phi_{c m} \right. \left. \left({\bm{y}}_{c m}\cdot \bm{n}\right) +\sin \Theta_{c m} \cos \Phi_{c m}\left({\bm{z}}_{c m} \cdot \bm{n}\right)\right] \\
& + \left.\gamma_{c m} E_p^{\prime} \frac{p \cos \Theta_n+q \cos \chi \cos \Theta_n+q \sin \chi \cos \Phi \sin \Theta_n}{E_p+E_q}\right\} - \frac{\gamma_{c m}}{1+\gamma_{c m}^2 v_{c m}^2} \frac{\cos \Theta_n}{p}\bigg\{\hat{p}_{c m}\left[\cos \Theta_{c m}\left({\bm{x}}_{c m} \cdot \bm{p}\right) \right.  \\
& \left. \left.+ \sin \Theta_{c m} \sin \Phi_{c m}\left({\bm{y}}_{c m} \cdot \bm{p}\right)\right]+\gamma_{c m} E_p^{\prime} \frac{p(p+ q \cos \chi)}{E_p+E_q}\right\}, 
\label{n_tilde_p_prime_avg.}
\end{aligned}
\end{equation}
where the different projections of Eq.~\eqref{n_tilde_p_prime_avg.} are mentioned in appendix~\ref{appendix_B}. 
\subsection{Radiative Process}
\begin{figure}
\centering
\begin{minipage}{0.26\linewidth}
\centering
\textbf{(a)}\\[-20pt]
\includegraphics[width=\linewidth]{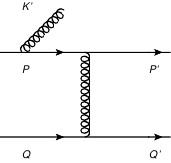}
\end{minipage}\hfill
\begin{minipage}{0.26\linewidth}
\centering
\textbf{(b)}\\ [8pt]
\includegraphics[width=\linewidth]{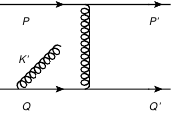}
\end{minipage}\hfill
\begin{minipage}{0.26\linewidth}
\centering
\textbf{(c)}\\[-20pt]
\includegraphics[width=\linewidth]{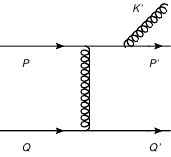}
\end{minipage}\hfill
\begin{minipage}{0.26\linewidth}
\centering
\textbf{(d)}\\ [8pt]
\includegraphics[width=\linewidth]{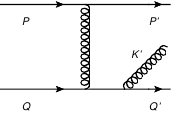}
\end{minipage}\hfill
\begin{minipage}{0.26\linewidth}
\centering
\textbf{(e)}\\[8pt]
\includegraphics[width=\linewidth]{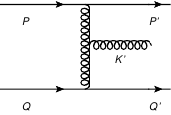}
\end{minipage}\hfill
\caption{Feynman diagrams for process $\mathrm{HQ}(P)+l(Q) \rightarrow$ $\mathrm{HQ}\left(P^{\prime}\right)+l\left(Q^{\prime}\right) + g(K^{\prime})$, showing the radiative processes of the $\mathrm{HQ}$ scattering with light quark and a soft gluon radiation.}
\label{radiative_process_graphs}
\end{figure}

In addition to elastic $2\rightarrow2$ scattering processes, gluon radiation in the medium is also possible, as illustrated in Fig.~\ref{radiative_process_graphs}. Such radiative processes play a significant role in estimating the transport properties of $\mathrm{{HQs}}$. An example of this is the inelastic scattering process: $\mathrm{HQ}(P)+l(Q) \rightarrow$ $\mathrm{HQ}\left(P^{\prime}\right)+l\left(Q^{\prime}\right) + g(K^{\prime})$, where $K^{\prime} = (E_{k^{\prime}},\bm{k}_{\perp}^{\prime},k_{z}^{\prime})$ denotes the four-momentum of the soft gluon emitted by the $\mathrm{HQ}$ in the final state of the interaction. The general expression for these $2\rightarrow3$ inelastic processes can be derived by substituting the two-body phase space factor and scattering amplitude with their respective three-body counterparts. This allows for the computation of the thermally averaged quantity$ \, \llangle \mathcal{F}(\bm{p})\rrangle$ for inelastic processes as
\ba
\llangle \mathcal{F}(\bm{p})\rrangle_{\mathrm{rad}} & \hspace{-0.1cm}=& \frac{1}{2 E_p g_{\mathrm{HQ}}}\int\hspace{-0.1cm} \frac{d^3 \bm{q}}{(2 \pi)^3\, 2E_q} \int\hspace{-0.1cm}\frac{d^3 \bm{q}^{\prime}}{(2 \pi)^3 \,2 E_{q^{\prime}}}\int\hspace{-0.1cm} \frac{d^3 \bm{p}^{\prime}}{(2 \pi)^3 \, 2E_{p^{\prime}}} 
 \int\hspace{-0.1cm}\frac{d^3 \bm{k}^{\prime}}{(2 \pi)^3 \,2 E_{k^{\prime}}} \sum\left|\mathcal{M}_{2 \rightarrow 3}\right|^2  \mathfrak{f}_k\left(E_q\right) \nn
&\times & \left(1 \pm \mathfrak{f}_k\left(E_{q^{\prime}}\right)\right) \left(1+\mathfrak{f}_g\left(E_{k^{\prime}}\right)\right) \Theta\left(E_p-E_{k^{\prime}}\right) \Theta\left(\tau-\tau_F\right) \mathcal{F}(\bm{p})\nn
&\times & (2 \pi)^4 \delta^{(4)} \left(P+Q-P^{\prime}-Q^{\prime}-K^{\prime}\right) \, .
\label{rad._process}
\ea 
Here, the scattering duration, which characterizes the time between successive interactions of the $\mathrm{HQs}$ with the medium constituents, is represented by $\tau$, and $\tau_F$ represents the gluon formation timescale. The $\Theta$ function imposes constraints on the dynamics of emitted gluons within the medium. Specifically, the condition $\Theta(E_p-E_{k'})$ ensures the emitted gluon's energy is less than the initial energy of the HQ. At the same time, $\Theta\left(\tau-\tau_F\right)$ puts the constraint that the formation time of gluon should be less than the interaction time of the $\mathrm{HQ}$ with the medium constituents and thus accounting for the Landau-Pomeranchuk-Migdal $(\mathrm{LPM})$ suppression effect~\cite{Wang:1994fx, Gyulassy:1993hr, Klein:1998du}. Also, $\mathfrak{f}_{g}(E_{k^\prime}) = 1/[\exp{(\beta E_{k^\prime})}-1]$ is the distribution of the radiated gluon where $\beta = 1/T$. It is assumed that the radiated gluon is in a thermally equilibrated state. The interaction term $\left|\mathcal{M}_{2 \rightarrow 3}\right|^2$ for the $2 \rightarrow 3$ radiative process can be expressed through the interaction term for the elastic process times the probability for soft gluon emission~\cite{Abir:2011jb} as follows,   
\begin{equation}
\left|\mathcal{M}_{2 \rightarrow 3}\right|^2=\left|\mathcal{M}_{2 \rightarrow 2}\right|^2 \times \frac{48\pi \alpha_{s}(T)}{{k}_{\perp}^{\prime 2}}\left(1+\frac{M_{\mathrm{HQ}}^2}{s} e^{2\zeta}\right)^{-2} \, ,
\end{equation}
where $\alpha_{s}(T)$ is the strong running coupling defined up to one-loop as 
\begin{equation}
\alpha_{s}(T) = \frac{6\pi}{11N_{c}-2N_{f}}\frac{1}{\ln (2\pi T/\Lambda_{\overline{\text{MS}}})} 
\end{equation}
having scale $\Lambda_{\overline{\text{MS}}}=176$ MeV~\cite{Haque:2014rua} for $N_{f}=3$. Additionally, $\zeta$ represents the rapidity of the emitted gluon, and the term $\left(1+M_{\mathrm{HQ}}^2 e^{2 \zeta}/s\right)^{-2}$ acts as a suppression factor for the HQ due to the dead-cone effect~\cite{Dokshitzer:2001zm, Abir:2011jb}. Consequently, from Eq.~\eqref{rad._process}, we obtain
\begin{equation}
\begin{aligned}
\llangle \mathcal{F}(\bm{p})\rrangle_{\mathrm{rad}}= & \llangle \mathcal{F}(\bm{p})\rrangle_{\mathrm{coll.}} \times \mathcal{I}(\bm{p}) \, ,
\end{aligned}
\end{equation}
where $\mathcal{I}(\bm{p})$ is given by
\begin{align}
\mathcal{I}\left(\bm{p}\right) &= \int \frac{d^3 k^{\prime}}{(2 \pi)^3 2 E_{k^{\prime}}} \frac{48 \pi \, \alpha_{s}(T)}{k_{\perp}^{\prime 2}}\left(1+\frac{M_{\mathrm{HQ}}^2}{s} e^{2 \zeta}\right)^{-2} \left(1+\mathfrak{f}_g\left(E_{k^{\prime}}\right)\right) \Theta\left(E_p-E_{k^{\prime}}\right) \Theta\left(\tau-\tau_F\right) \, .
\label{Int_I_k}
\end{align}
\begin{figure}
\centering
\includegraphics[width=8cm, height=6cm]{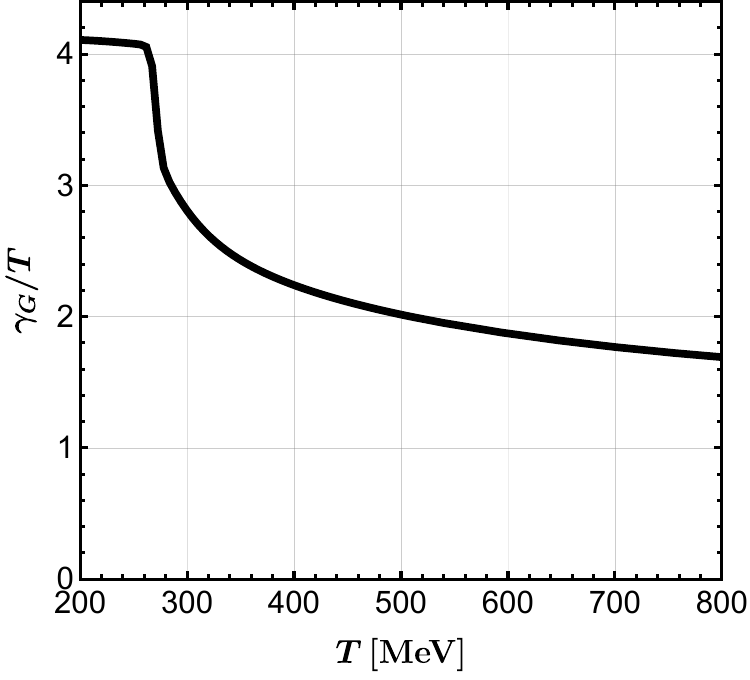}
\caption{Temperature dependence of scaled Gribov mass parameter $\gamma_{G}/T$ determined via lattice (thermodynamics) data~\cite{Borsanyi:2012ve}.}
\label{gammaG_variation}
\end{figure}
The Eq.~\eqref{Int_I_k} can be simplified further by noting the relations between the energy and transverse momentum of the emitted gluon with the rapidity variable as follows:
\begin{align}
E_{k^{\prime}} = k_{\perp}^{\prime} \cosh\zeta \, , \quad \quad k_{z}^{\prime} = k_{\perp}^{\prime} \sinh\zeta \, ,
\end{align}
with $d^{3}k^{\prime} = d^{2}k_{\perp}^{\prime}dk_{z}^{\prime} = 2\pi k_{\perp}^{\prime 2}dk_{\perp}^{\prime}\cosh\zeta \,d\zeta$. The interaction time $\tau$ is related to the interaction rate $\Gamma = 2.26 \alpha_{s}T$ and the $\Theta(\tau-\tau_{F})$ impose the constraint $\tau = \Gamma^{-1} > \tau_{F} = (\cosh\zeta)/(k_{\perp}^{\prime})$ 
which puts a lower cut-off on $k_{\perp}^{\prime}$. While the other theta function i.e. $\Theta(E_{p}-E_{k^{\prime}})$ constraint the upper limit of $(k_{\perp}^{\prime})_{\text{max}} = (E_{p})/\cosh\zeta$. Thus in the limiting case ($E_{k^{\prime}} \ll T$), the integral $\mathcal{I}(\bm{p})$ becomes
\begin{align}
\mathcal{I}\left(\bm{p}\right) &=  \, \frac{6}{\pi} \alpha_{s} T \int_{\Gamma \cosh \zeta}^{E_p / \cosh \zeta} d k_{\perp}^{\prime} \int_{-\zeta_{1}}^{\zeta_{1}} d \zeta  \left(1+\frac{M_{\mathrm{HQ}}^2}{s} e^{2 \zeta}\right)^{-2} \frac{1}{k_{\perp}^{\prime \,2} \cosh \zeta} \, ,
\end{align}
where we have used the rapidity integration limits as $\zeta_{1}=20$. 
\section{RESULTS AND DISCUSSION}\label{section_3}
To perform the numerical computation of various transport coefficients of $\mathrm{HQs}$, we need to fix first the Gribov mass parameter $\gamma_{G}$. This has been achieved via determining the equilibrium thermodynamics quantities of Gribov plasma~\cite{Jaiswal:2020qmj} and doing the matching of the said quantities with the scaled trace anomaly results obtained through lattice~\cite{Borsanyi:2012ve}. Fig.~\ref{gammaG_variation} shows the dependence of the scaled Gribov parameter $\gamma_{G}/T$ with temperature $T$, which is in accord with the theoretical calculations in the high-temperature limit~\cite{Fukushima:2013xsa}. This dependence has been utilized in the transport coefficient through the interaction elements presented in Appendix~\ref{appendix_A}. We now discuss the numerically estimated results for the drag and diffusion coefficients, energy loss, and nuclear modification factor of the charm quark, particularly with a mass of $1.3~$GeV in an anisotropic medium, in separate subsections.    
\subsection{Transport coefficients in anisotropic medium}
\begin{figure}
\centering
\begin{minipage}{0.47\linewidth}
\centering
\textbf{(a)}\\
\includegraphics[width=\linewidth]{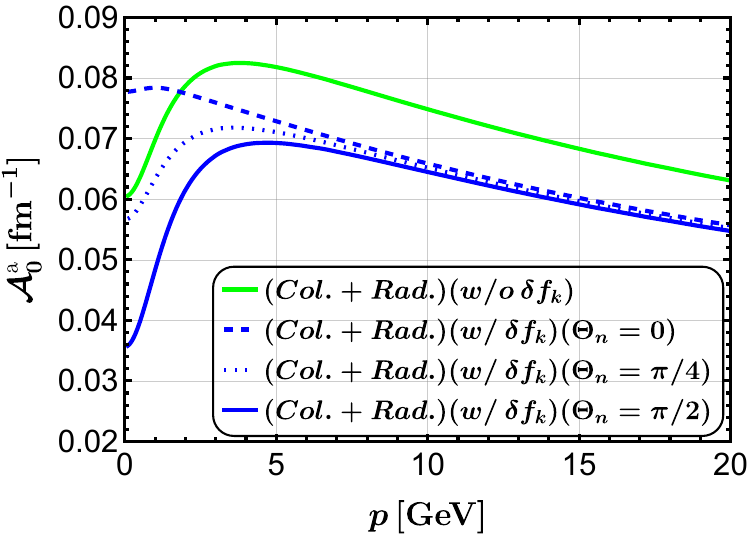}\label{drag_coefficient_total}
\end{minipage}\hfill
\begin{minipage}{0.47\linewidth}
\centering
\textbf{(b)}\\ 
\includegraphics[width=\linewidth]{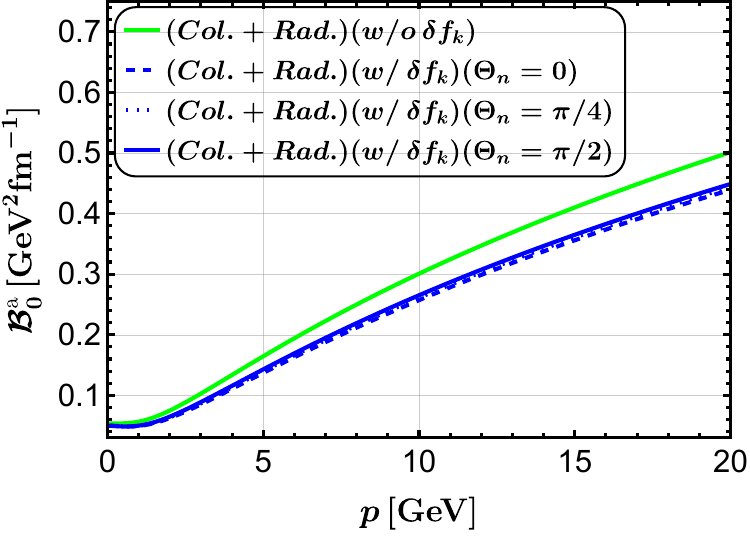}\label{diffusion_coefficient_total}
\end{minipage}\hfill
\caption{\textit{Left panel:} Momentum variation of total drag coefficient in an anisotropic medium for different directional anisotropies. \textit{Right panel:} Total $\mathrm{HQ}$ diffusion coefficient $\mathcal{B}_{0}^{\text{(a)}}$ variation with momentum. Both drag and diffusion coefficients are calculated for $T = 360$ MeV}
\label{drag_+_diff.}
\end{figure}

To consider the anisotropy effects, we have considered the weak anisotropy scenario where $\xi$ can take $0.2, 0.3,0.4$ values only, which have been obtained through the hydrodynamic simulation approach~\cite{Epelbaum:2013ekf, Strickland:2014eua}. In Fig.~\ref{drag_+_diff.}\,(a), we have plotted the drag coefficient in an anisotropic medium for a particular anisotropy parameter $\xi = 0.3$ and for various directional anisotropies accounting for both the collisional and radiative contributions combined. It can be observed that the $\mathrm{HQs}$ suffer a lesser hindrance in an anisotropic medium compared to an isotropic one. Also, the presence of hindrance will become much less for the lower momentum regime as the directional anisotropy increases, as can be seen in Fig.~\ref{drag_+_diff.}\,(a). The reduction of the drag coefficient at large $p_T$ should be understood, relative to the isotropic case. While the anisotropic medium yields a smaller drag coefficient, the relative influence of momentum-space anisotropy decreases with increasing $p_T$, as expected for weak anisotropy. We emphasize that the above comparison is performed at a fixed hard scale. In the absence of a matching condition, changing $\xi$ modifies not only the angular structure of the distribution but also the thermodynamic densities of the medium. Therefore, the fixed-hard-scale results should not be interpreted as a comparison at a strictly fixed number, energy, or entropy density. \\
A possible alternative is to impose a matching condition between the isotropic and anisotropic systems, for example, at fixed number, energy, or entropy density. Among these choices, energy-density matching is particularly natural because it compares systems with the same overall energy content while retaining momentum-space anisotropy. For an illustrative estimate, we use the standard massless conformal form of the RS ansatz, for which~\cite{Martinez:2010sc, Strickland:2014pga}
\begin{equation}
\varepsilon_{\xi}(\Lambda_{\xi}) = {\cal R}(\xi)\varepsilon_0(\Lambda_{\xi}),
\qquad {\cal R}(\xi) = \frac{1}{2} \left[ \frac{1}{1+\xi} + \frac{\tan^{-1}\sqrt{\xi}}{\sqrt{\xi}} \right], \qquad \xi>0 .
\end{equation}
The energy-density matching condition gives $\Lambda_{\xi} = T\,{\cal R}(\xi)^{-1/4}. $ For $\xi=0.3$, one obtains ${\cal R}(0.3)\simeq0.842$ and hence $\Lambda_{\xi}\simeq1.044T$. At $T=360~{\rm MeV}$, this corresponds to $\Lambda_{\xi}\simeq376~{\rm MeV}$. Thus, for the weak anisotropy used in Fig.~3, energy-density matching leads to a moderate shift of the hard scale. \\
This also clarifies the high-momentum behavior shown in Fig.~\ref{drag_+_diff.} (a). In the fixed-hard-scale setup, the number density scales as $n_\xi(T)=n_0(T)/\sqrt{1+\xi}$; for $\xi=0.3$, this gives $n_\xi(T)\simeq0.877\,n_0(T)$, corresponding to a reduction of about $12.3\%$. Similarly, the energy-density factor ${\cal R}(0.3)\simeq0.842$ corresponds to a reduction of about $15.8\%$ at the same hard scale. Therefore, the approximately $15\%$ reduction observed in the high-momentum region should not be attributed solely to the angular deformation of the distribution at fixed thermodynamic densities. It may also reflect the accompanying change in the thermodynamic densities in the fixed-hard-scale setup. The precise numerical size of the anisotropic correction can therefore depend on the chosen matching prescription. The momentum dependence of diffusion coefficient component $\mathcal{B}_{0}^{\text{(a)}} = \mathcal{B}_{0} + \delta \mathcal{B}_{0}$ is shown in Fig.~\ref{drag_+_diff.}\,(b) where anisotropy lowers the $\mathcal{B}_{0}^{\text{(a)}}$ component magnitude for higher momentum values significantly compared to lower momentum values. The directional anisotropy does not significantly affect momentum diffusion $\mathcal{B}_{0}^{\text{(a)}}$. \\
\begin{figure}
\centering
\begin{minipage}{0.33\linewidth}
\centering
\textbf{(a)}\\
\includegraphics[width=\linewidth]{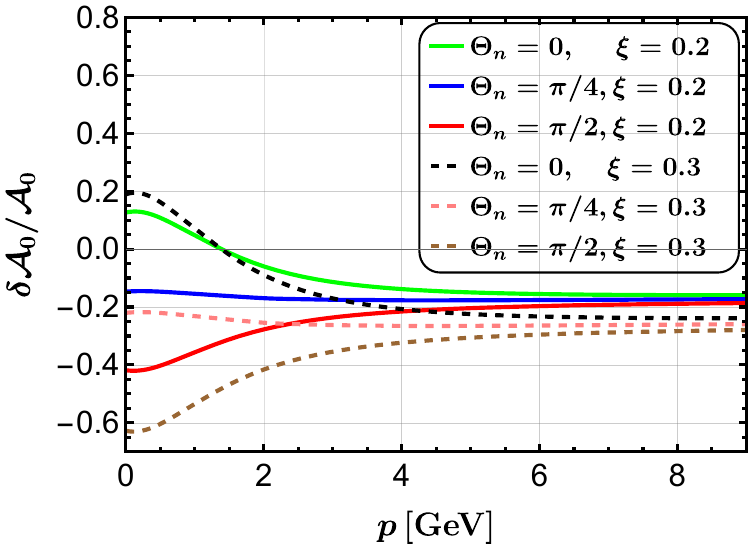}\label{DA0}
\end{minipage}\hfill
\begin{minipage}{0.33\linewidth}
\centering
\textbf{(b)}\\
\includegraphics[width=\linewidth]{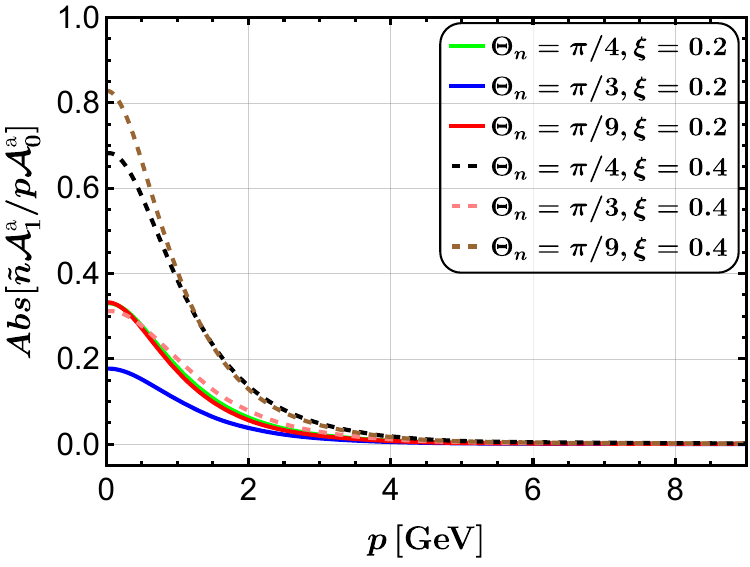}\label{Abs_NTA1_RHIC}
\end{minipage}\hfill
\begin{minipage}{0.33\linewidth}
\centering
\textbf{(c)}\\
\includegraphics[width=\linewidth]{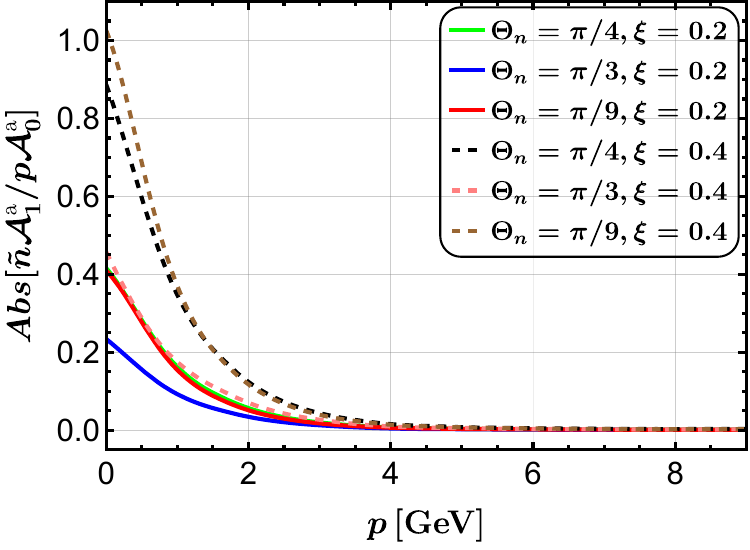}\label{Abs_NTA1_LHC}
\end{minipage}
\caption{\textit{Left panel:} Momentum dependence for the ratio of anisotropic correction to the isotropic drag coefficient for $T = 360$ MeV. \textit{Middle panel:} The relative significance of $\mathcal{A}_{1}^{\text{(a)}}$ with $\mathcal{A}_{0}^{\text{(a)}}$ for $T = 360$ MeV. \textit{Right panel:} The relative significance of $\mathcal{A}_{1}^{\text{(a)}}$ with $\mathcal{A}_{0}^{\text{(a)}}$ for $T = 480$ MeV.}
\label{DA0_and_Abs_NTA1}
\end{figure}
In Fig.~\ref{DA0_and_Abs_NTA1}\,(a), we have shown how the ratio of anisotropic correction to isotropic contribution varies with the $\mathrm{HQ}$ momentum. It can be seen that the said ratio depends strongly on the anisotropy strength and directional anisotropy for lower momentum values compared to the high-momentum regime. The anisotropic correction with respect to the isotropic contribution to the drag coefficient reduces as the anisotropy strength increases, along with the increase in the angle between the anisotropy direction $(\Theta_{n})$ and the $\mathrm{HQ}$ motion direction. The other drag coefficient, which solely arises due to anisotropy effects, has been plotted in Figs.~\ref{DA0_and_Abs_NTA1}\,(b) and~\ref{DA0_and_Abs_NTA1}\,(c) for RHIC and LHC energies, respectively, which shows a diminishing dependence on high-momentum values. Again, the anisotropy strength plays a key role in the additional drag component, which is more significant at LHC energies compared to RHIC energies. Note that here we have added both the collisional and the radiative contributions by considering the nonperturbative effects via the Gribov-Zwanziger propagator, contrary to the earlier perturbative results where only the collisional contribution has been included in~\cite{Kumar:2021goi} and radiative processes effects have been considered in~\cite{Prakash:2023wbs}.   \\
The anisotropic corrections for the components $\mathcal{B}_{0}^{\text{(a)}} = \mathcal{B}_{0} + \delta \mathcal{B}_{0} $ and $\mathcal{B}_{1}^{\text{(a)}} = \mathcal{B}_{1} + \delta \mathcal{B}_{1} $  with respect to isotropic $\mathrm{HQ}$ diffusion coefficient has been shown in Figs.~\ref{DB0_and_DB1}\,(a) and~\ref{DB0_and_DB1}\,(b), respectively at $T= 360$ MeV. For both diffusion components, the anisotropy parameter $\xi$ has a strong dependence and thus lowers the diffusion components significantly. Also for the component $\mathcal{B}_{0}^{\text{(a)}}$ got an extra anisotropic contribution compared to $\mathcal{B}_{1}^{\text{(a)}}$ as can be seen from the Eq.~\eqref{diffusion_components}. It has been observed that the directional anisotropy is more prominent for the intermediate and the higher momentum regime for $\delta\mathcal{B}_{0}$ component compared to $\delta\mathcal{B}_{1}$ component. \\ 
\begin{figure}
\centering
\begin{minipage}{0.45\linewidth}
\centering
\textbf{(a)}\\
\includegraphics[width=\linewidth]{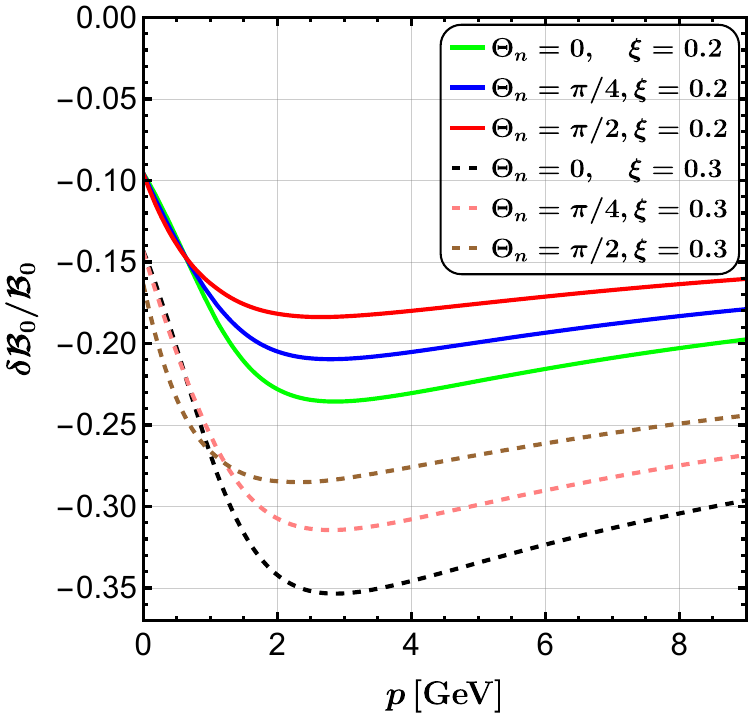}\label{DB10by_B0.pdf}
\end{minipage}\hfill
\begin{minipage}{0.45\linewidth}
\centering
\textbf{(b)}\\
\includegraphics[width=\linewidth]{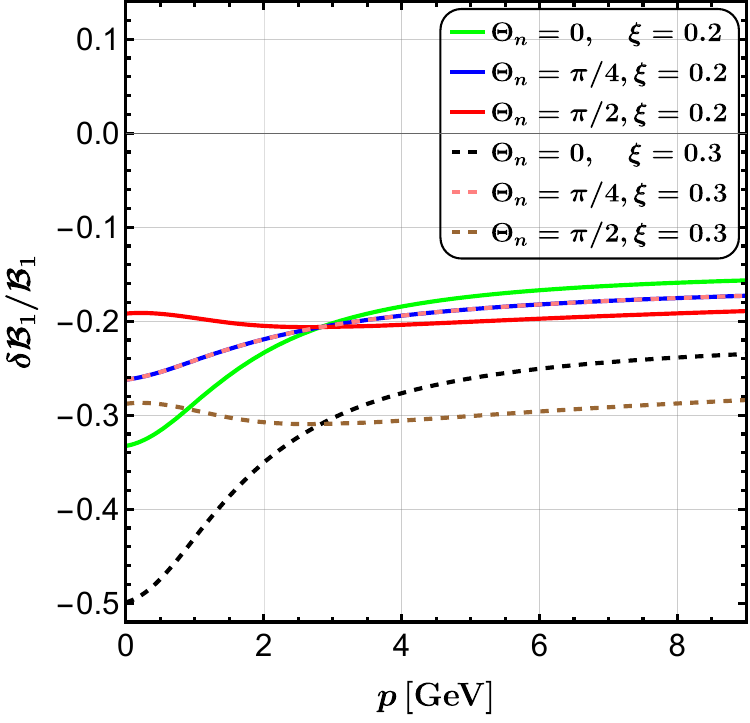}\label{DB1_by_B1.pdf}
\end{minipage}\hfill
\caption{\textit{Left panel:} Momentum variation of anisotropic correction to isotropic diffusion coefficient for $\mathcal{B}_{0}$. \textit{Right panel:} Momentum variation of anisotropic correction to isotropic diffusion coefficient for $\mathcal{B}_{1}$.}
\label{DB0_and_DB1}
\end{figure}
\begin{figure}
\centering
\begin{minipage}{0.32\linewidth}
\centering
\textbf{(a)}\\
\includegraphics[width=\linewidth]{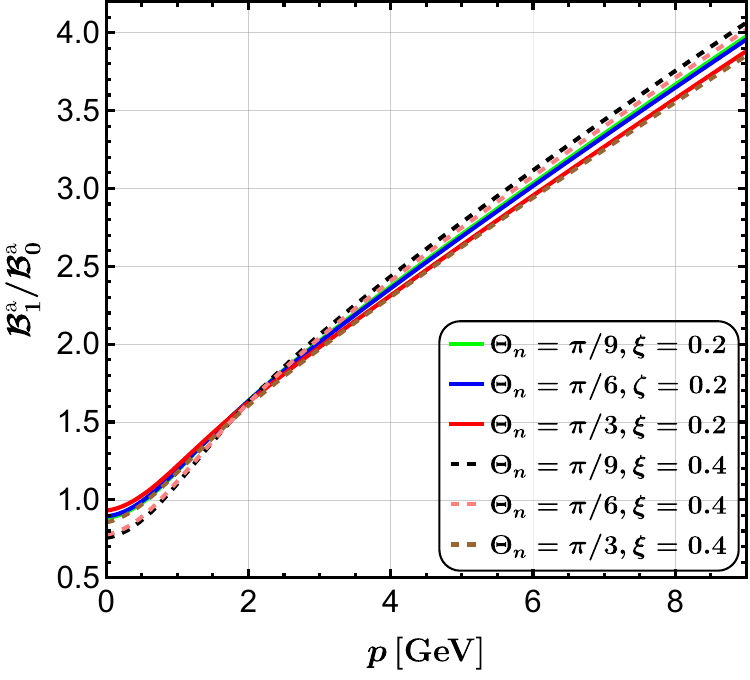}\label{B1_by_B0.pdf}
\end{minipage}\hfill
\begin{minipage}{0.32\linewidth}
\centering
\textbf{(b)}\\
\includegraphics[width=\linewidth]{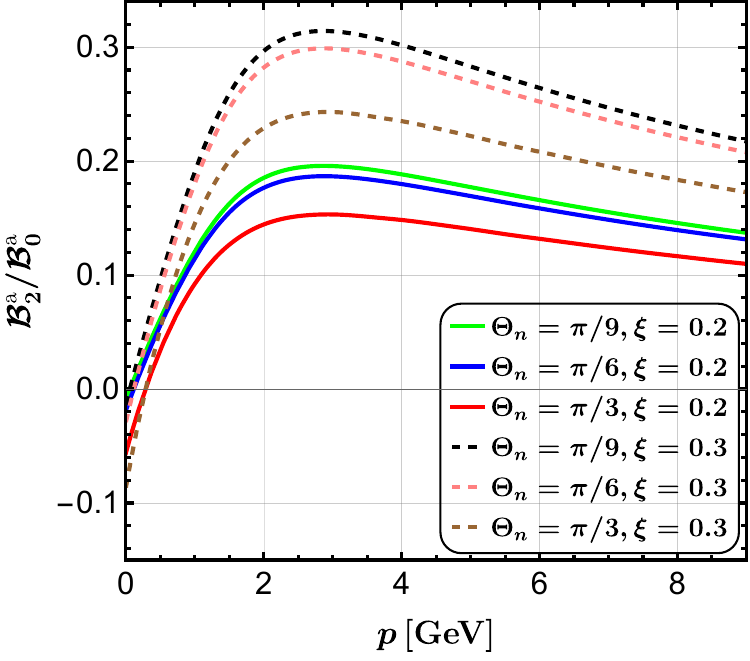}\label{B2_by_B0.pdf}
\end{minipage}\hfill
\begin{minipage}{0.32\linewidth}
\centering
\textbf{(c)}\\
\includegraphics[width=\linewidth]{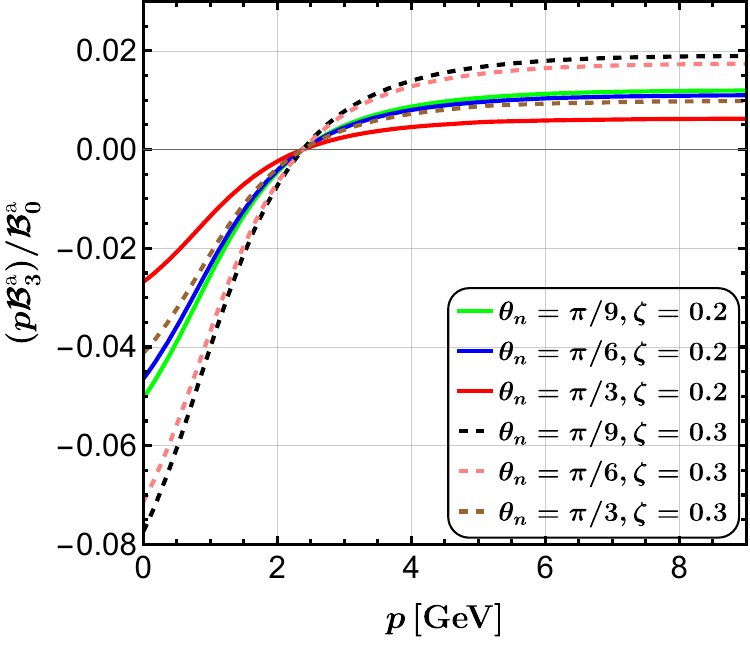}\label{pB3_by_B0.pdf}
\end{minipage}\hfill
\caption{Momentum variation of the ratio of different diffusion coefficient components, namely $\mathcal{B}_{1}$\textit{(Left panel)}, $\mathcal{B}_{2}$ \textit{(Middle panel)}, and $\mathcal{B}_{3}$ \textit{(Right panel)} with $\mathcal{B}_{0}$ at $T = 360$ MeV. }
\label{B1_and_B2_and_pB3}
\end{figure}
In Fig.~\ref{B1_and_B2_and_pB3}\,(a), we have plotted the relative correction to the $\mathrm{HQ}$ diffusion coefficient namely $\mathcal{B}_{1}^{\text{(a)}}$. It has been found that the diffusion component $\mathcal{B}_{1}^{\text{(a)}}$ is roughly same with $\mathcal{B}_{0}^{\text{(a)}}$ for low-momentum regime. At the same time, substantial deviations have been observed across different directional anisotropies in the high-momentum regime. The other two diffusion components, $\mathcal{B}_{2}^{\text{(a)}}$ and $\mathcal{B}_{3}^{\text{(a)}}$, which arises only due to anisotropy effects have been plotted relatively with respect to $\mathcal{B}_{0}^{\text{(a)}}$ in Figs.~\ref{B1_and_B2_and_pB3}\,(b) and \ref{B1_and_B2_and_pB3}\,(c) respectively which shows that the anisotropy parameter $\xi$ and directional anisotropy parameter $\Theta_{n}$ affects the $\mathcal{B}_{2}^{\text{(a)}}$ component more compared to $\mathcal{B}_{3}^{\text{(a)}}$. It is worth noting that even in the limit $p\to 0$, the transport coefficients may retain a dependence on the anisotropy direction $(\Theta_{n})$. This arises because the background medium itself is anisotropic, and the distribution function of the medium particles explicitly contains the preferred direction ${\bf n}$.
\subsection{Energy loss of $\mathrm{HQs}$ in an anisotropic medium}
\begin{figure}
\centering
\begin{minipage}{0.48\linewidth}
\centering
\textbf{(a)}\\
\includegraphics[width=\linewidth]{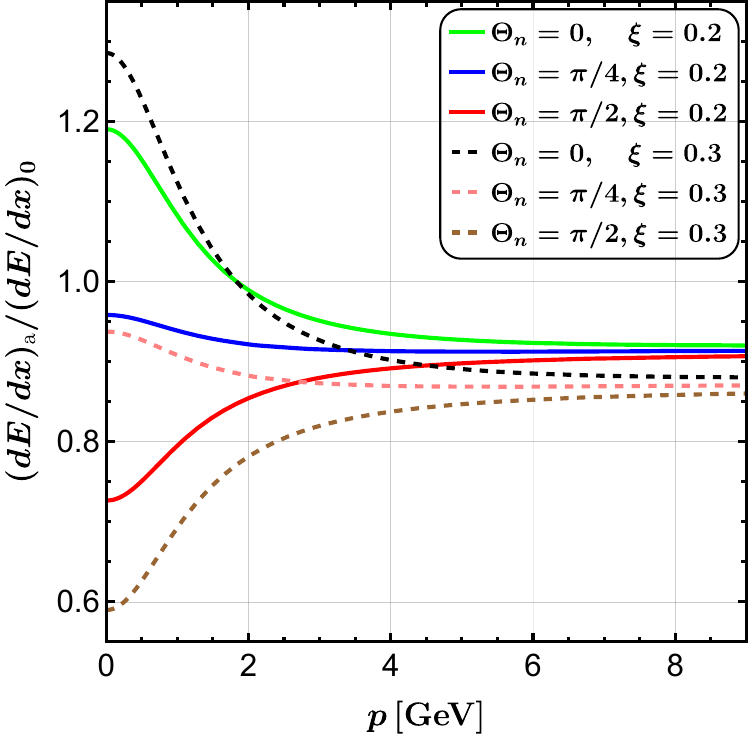}\label{Energy_Loss_RHIC}
\end{minipage}\hfill
\begin{minipage}{0.48\linewidth}
\centering
\textbf{(b)}\\
\includegraphics[width=\linewidth]{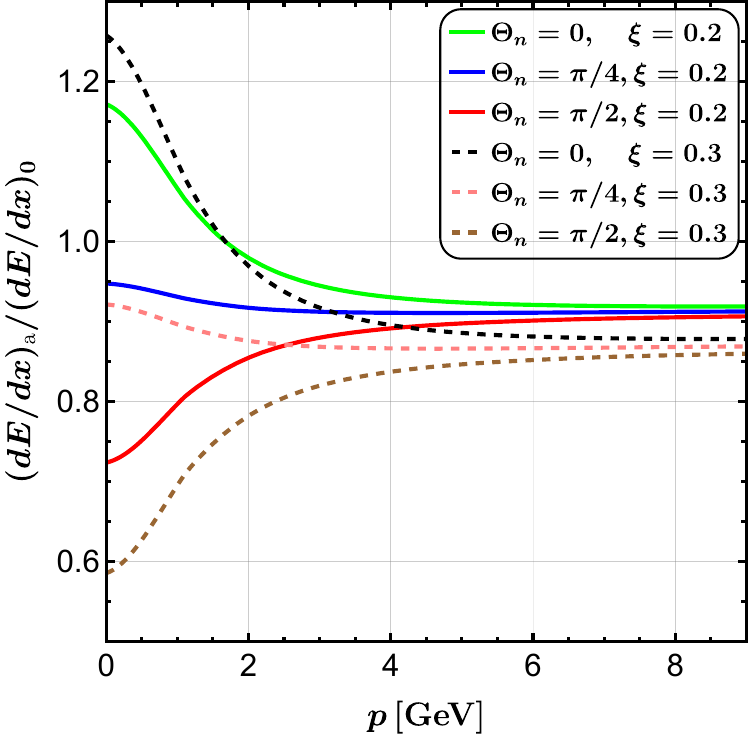}\label{Energy_Loss_LHC}
\end{minipage}\hfill
\caption{Relative significance of the $\mathrm{HQ}$ energy loss for different weak anisotropy strengths at RHIC (\textit{Left panel}) and LHC energy (\textit{Right panel}).}
\label{Energy_Loss}
\end{figure}
As $\mathrm{HQs}$ traverses through the anisotropic $\mathrm{QCD}$ medium, the energy dissipation of $\mathrm{HQs}$ occurs mainly through collisional and radiative processes, and this differential energy loss can be described through the $\mathrm{HQ}$ drag coefficient in the following manner~\cite{GolamMustafa:1997id}
\begin{equation}
\begin{aligned}
\left( \frac{dE}{dx} \right)_{\text{a}} = \mathcal{A}_{0}^{\text{(a)}} \, (p^{2},T).
\end{aligned}
\end{equation}
The other drag component, namely $\mathcal{A}_{1}^{\text{(a)}} \, (p^{2},{T})$ will not contribute to the present analysis since we are considering the energy loss in the direction of $\mathrm{HQ}$ initial momentum. Thus, the anisotropy contribution comes through $\delta \mathcal{A}_{0}$ only as $p \cdot \tilde{n} = 0$. In Fig.~\ref{Energy_Loss}, we have plotted the momentum dependence of the ratio of relative energy loss, i.e., anisotropic correction to energy loss $(dE/dx)_{\text{a}}$ with the isotropic contribution $(dE/dx)_{0}$ for the RHIC and LHC energies. It can be observed that the direction of anisotropy plays a key role, along with the strength of the anisotropy, since the said ratio of energy loss gets suppressed with an increase in the anisotropy strength. Additionally, directional anisotropy contributes more to lower-momentum regimes than to higher-momentum regimes, and it again suppresses the energy-loss ratio. The same observations also hold for LHC energies, with even greater clarity.
\subsection{Nuclear Modification factor, $\mathrm{R_{AA}}$}
\begin{figure}
\centering
\begin{minipage}{0.48\linewidth}
\centering
\textbf{(a)}\\
\includegraphics[width=\linewidth]{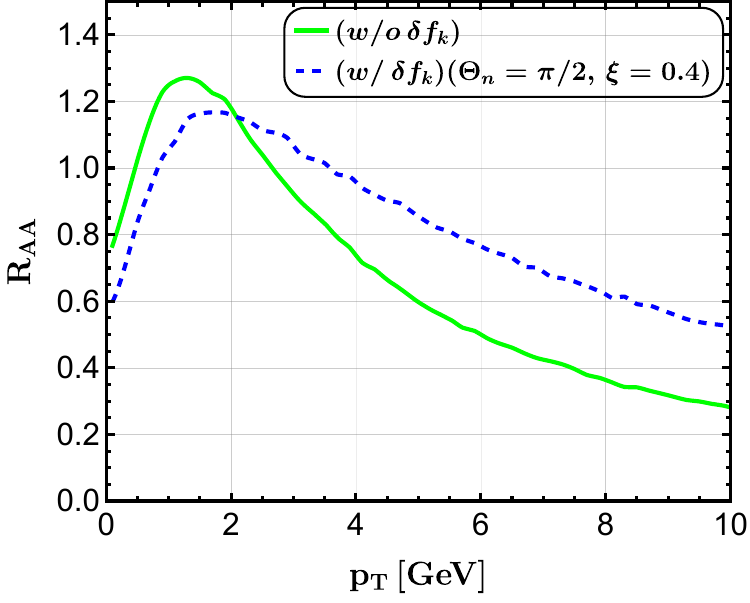}\label{RAA_RHIC}
\end{minipage}\hfill
\begin{minipage}{0.48\linewidth}
\centering
\textbf{(b)}\\
\includegraphics[width=\linewidth]{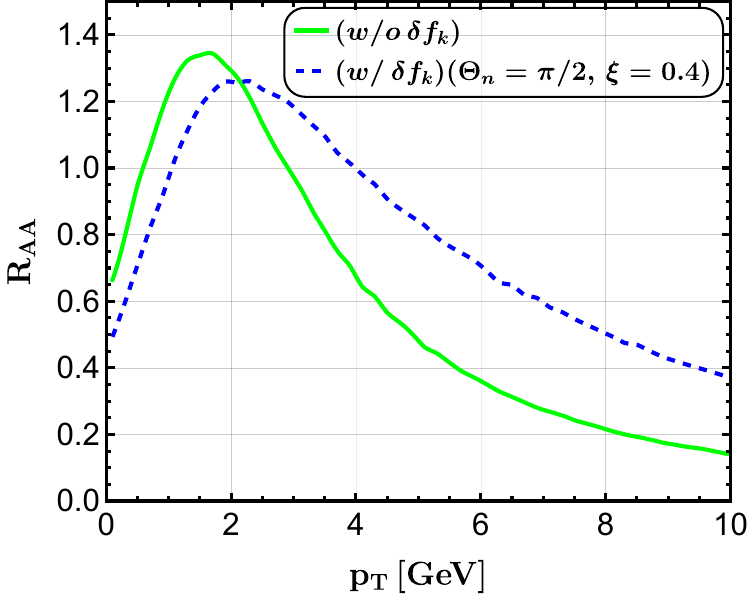}\label{RAA_LHC}
\end{minipage}\hfill
\caption{Nuclear modification factor ($\mathrm{R_{AA}}$) is plotted as a function of transverse momentum ($\mathrm{p_T}$) for a QGP evolution time $\tau_f = 5$ fm/c. Results are shown at two temperatures: $T = 360$ MeV \textit{(Left panel)} and  $T = 480$ MeV \textit{(Right panel)}, considering collisional and radiation combined processes for charm and bottom quarks denoted by $c$ and $b$ respectively.}
\label{RAA}
\end{figure}

We have investigated how momentum anisotropy influences the $\mathrm{R_{AA}}$ of $\mathrm{HQs}$ in anisotropic media.
We have defined the \(\mathrm{R_{AA}}({p_T})\) as follows,
\begin{align}
 \mathrm{R_{AA}}({p_T)}=\frac{f_{\tau_f} ({p_T} )}{f_{\tau_0} ({p_T})},
 \end{align}
where \( f_{\tau_f}({p_T}) \) represents the momentum distribution of charm quarks after evolving for a time \( \tau_f = 5 \) fm/c, and \( f_{\tau_0}({p_T}) \) corresponds to the initial momentum distribution of charm quarks. The initial distribution is taken from Fixed Order + Next-to-Leading Log (FONLL) calculations, which accurately describe the production spectra of D-mesons in proton-proton collisions via fragmentation processes~\cite{cacciari2005qcd, Cacciari:2012ny}. The initial momentum spectrum is parameterized as
\begin{equation}
\frac{dN}{d^2p{_T}} = \frac{x_0}{(x_1+{p_T})^{x_2}},
\end{equation}
with the parameters \( x_0 = 6.365480 \times 10^8 \), \( x_1 = 9.0 \), and \( x_2 = 10.27890 \). A deviation of \(\mathrm{R_{AA}}(p_T) \) from unity indicates that charm quarks interact with the medium, resulting in significant alterations to their momentum spectra. These interactions reflect the medium's influence on the dynamics of charm quarks. The standard method for studying the momentum evolution of the $\mathrm{HQ}$ in a medium involves solving the Fokker-Planck equation using a stochastic Langevin dynamics approach.
The corresponding Langevin equation of motion for the $\mathrm{HQ}$ is given by~\cite{Moore:2004tg}
\begin{equation}
dx_i=\frac{p_i}{E} dt, \quad \quad \quad dp_i=-\mathcal{A}p_i\, dt+C_{ij}\rho_j\sqrt{dt},
\end{equation}
Here, $dx_i$ and $dp_i$ represent the respective changes in position and momentum over each time interval $dt$. The drag coefficient is denoted by $\mathcal{A}$. $C_{ij}$ represents the covariance matrix, and $\rho_j$ is a Gaussian-distributed random variable with $\langle\rho_i\rho_j\rangle=\delta_{ij}$ and $\langle\rho_i\rangle=0$. The matrix $C_{ij}$ is expressed as
\begin{align}
C_{ij} = \sqrt{2\mathcal{B}_0}\left(\delta_{ij}-\frac{p_i p_j}{p^2}\right) + \sqrt{2\mathcal{B}_1} \frac{p_i p_j}{p^2}.
\end{align}
In the limit $\mathcal{B}_0=\mathcal{B}_1=\mathcal{B}$, we have $C_{ij}=\sqrt{2\mathcal{B}}\delta_{ij}$. 
While this isotropic approximation is rigorously valid only in the static limit ($p\rightarrow 0$), it remains widely utilized for modeling $\mathrm{HQ}$ motion at finite momentum in the QGP medium~\cite{Rapp:2018qla, Moore:2004tg,vanHees:2005wb}.\\
We have calculated $\mathrm{R_{AA}}$ to assess the effects of the Gribov gluon plasma in the presence of an anisotropic medium. The results are shown in Fig.~\ref{RAA} for two temperature values: $360$ MeV (\textit{Left panel}), corresponding to RHIC energies, and $480$ MeV, corresponding to LHC energies (\textit{Right panel}) for both charm and bottom quarks. A static medium is assumed with a lifetime of $\tau_f$=5 fm/c as the characteristic lifetime of the QGP typically expected in high-energy nuclear collisions at RHIC and LHC energies. The impact of momentum anisotropy is evaluated for an anisotropy angle $\Theta_{n} = \pi/2$ and an anisotropy strength $\xi = 0.4$. The effects of collisional and radiative energy loss mechanisms are incorporated for both isotropic and anisotropic media.  
The results indicate that the suppression of $\mathrm{R_{AA}}$ is more pronounced in the isotropic medium than in the anisotropic medium at higher $\mathrm{p_T}$. At low momentum,  $\mathrm{R_{AA}}$ rises above unity as a result of $\mathrm{HQ}$ number conservation. Interactions of $\mathrm{HQs}$ with the thermal bath cause high-momentum $\mathrm{HQs}$ to lose energy and shift to lower momentum, leading to an enhanced population in the low-momentum region. Additionally, the results exhibit a clear mass hierarchy in both isotropic and anisotropic media at $T = 360$ MeV and $T = 480$ MeV. For a given $\mathrm{p_T}$, bottom quarks show larger $\mathrm{R_{AA}}$ than charm quarks systematically, reflecting their reduced in-medium energy loss due to the larger mass. For example at $\mathrm{p_T} \approx 5$ GeV $\mathrm{R_{AA}}^{b}$ exceeds $\mathrm{R_{AA}}^{c}$ by about $10-25 \%$ depending on temperature and isotropy, with the largest difference observed in the isotropic medium at $T = 480$ MeV. The presence of momentum anisotropy $(\Theta_{n} = \pi/2, \xi = 0.4)$ mitigates the overall suppression for both flavors. However, the ordering $\mathrm{R_{AA}}^{b} > \mathrm{R_{AA}}^{c}$  remains intact across the entire $\mathrm{p_T}$ range and both temperatures, in agreement with the expected mass ordering of $\mathrm{HQ}$ energy loss in the QGP medium. Note that the present calculation is performed within a static medium approximation, and therefore does not include radial flow effects, which may influence the detailed shape of the nuclear modification factor. The results presented here should be regarded as qualitative estimates of anisotropy-induced modifications of $\mathrm{HQs}$ energy loss.
\section{CONCLUSION}\label{section_4}
To summarize, in this work, we have investigated the behavior of $\mathrm{HQs}$ as they undergo both elastic and radiative energy loss. At the same time, they interact with the medium constituents in an anisotropic QGP medium. We have studied the dynamics of $\mathrm{HQs}$ for different anisotropy strengths and various directional anisotropies present in the hot $\mathrm{QCD}$ environment. The study examines both elastic and inelastic interactions between $\mathrm{HQs}$ and the medium by evaluating the corresponding drag and diffusion coefficients, employing the Fokker-Planck formalism as the theoretical framework. The medium interactions between the $\mathrm{HQs}$ and the medium constituents have been modeled using the Gribov propagator to incorporate nonperturbative effects. The effects of momentum anisotropy have been included through deviations in the parton equilibrium distribution function, considered in the weak-anisotropy limit. In an anisotropic medium, the drag and diffusion processes split into multiple components due to the directional dependence introduced by the anisotropy. Specifically, the drag coefficients are decomposed into two distinct components, and the diffusion coefficients are decomposed into four distinct coefficients. The matching between the thermodynamics of the Gribov plasma and the pure gauge lattice results has been performed to estimate the temperature dependence of the Gribov mass parameter. This dependence has been utilized to estimate the anisotropic corrections of the drag and diffusion coefficients. \\
It has been deduced that in the presence of weak momentum anisotropy, the QCD medium exerts less hindrance to $\mathrm{HQs}$  and, further, the hindrance becomes less as the directional anisotropy increases in the medium. Also, momentum and directional anisotropy are more dominant in the lower momentum regime than in higher momentum values for both drag coefficient components. For various diffusion coefficients, the degree of momentum anisotropy and the angular dependence between the anisotropy vector and the $\mathrm{HQ}$ momentum direction significantly influence their estimation. The drag and diffusion coefficients obtained via the Fokker-Planck equation are utilized to calculate the energy loss and $\mathrm{R_{AA}}$ of the $\mathrm{HQs}$, and it has been found that the anisotropy strength and directional anisotropy suppress the energy loss for both RHIC and LHC energies. Also, $\mathrm{R_{AA}}$ is more pronounced at higher transverse momentum $\mathrm{p_T}$ in the isotropic medium compared to the anisotropic medium. Also note that, to study the impact of momentum-space anisotropy in a controlled, simplified setup, we employ fixed values for the hard scale and evolution time.
We emphasize that the present results should be understood within a weak-anisotropy framework, where only linear terms in $\xi$ are retained. A complete treatment of strongly anisotropic early-time dynamics, coherent Glasma fields, or a full hydrodynamic evolution of the shear-stress tensor lies beyond the scope of the present work and will be considered in future studies. Therefore, the phenomenological results should be interpreted as qualitative illustrations of anisotropy effects rather than precise quantitative predictions. Finally, we note that the numerical results presented here are obtained in a fixed-hard-scale setup without imposing thermodynamic matching; a fully matched calculation, for example, based on energy-density matching, may modify the precise numerical size of the anisotropic corrections and will be useful for future quantitative studies.\\
The phenomenological analysis presented in this work aims to provide a qualitative illustration of the effects of momentum-space anisotropy on $\mathrm{HQ}$ transport in a simplified medium. A fully realistic description would require coupling the transport coefficients derived here to a space-time dependent hydrodynamic evolution of the quark-gluon plasma. Such an implementation is beyond the scope of the present study and will be pursued in future work. It will be interesting to study the elliptic flow coefficient ($\mathrm{v_2}$) of $\mathrm{HQs}$ in an anisotropic background, where anisotropy may play a significant role. We will address this aspect in a subsequent publication. As a further extension, we aim to investigate the role of bulk viscosity in the $\mathrm{HQ}$ transport coefficients and their implications for experimental signatures such as $\mathrm{R_{AA}}$, and $\mathrm{v_2}$. Such an analysis could deepen our understanding of QGP transport behavior in extreme environments. To this end, we plan to incorporate bulk viscous effects through nonequilibrium modifications to the distribution functions of medium partons within the current theoretical framework~\cite{Thakur:2020ifi}.  
\begin{figure}
\centering
\begin{minipage}{0.15\linewidth}
\centering
\textbf{(a)}\\[0pt]
\includegraphics[width=\linewidth]{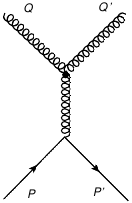}
\end{minipage}\hfill
\begin{minipage}{0.25\linewidth}
\centering
\textbf{(b)}\\ [0pt]
\includegraphics[width=\linewidth]{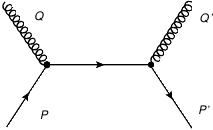}
\end{minipage}\hfill
\begin{minipage}{0.25\linewidth}
\centering
\textbf{(c)}\\[0pt]
\includegraphics[width=\linewidth]{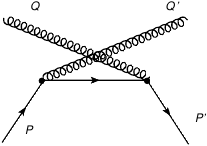}
\end{minipage}\hfill
\begin{minipage}{0.15\linewidth}
\centering
\textbf{(d)}\\ [0pt]
\includegraphics[width=\linewidth]{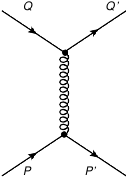}
\end{minipage}\hfill
\caption{Feynman graphs for the $\mathrm{HQ}$ elastic scattering interaction with (a) gluon ($t-$ type exchange), (b) gluon ($s-$ type exchange), (c) gluon ($u-$ type exchange), (d) light quark/antiquark ($t-$ type exchange).}%
\label{collision_feynman_graphs}%
\end{figure}
\begin{figure*}
\appendix
\section{SCATTERING ELEMENTS FOR ELASTIC PROCESSES}\label{appendix_A}
The leading-order Feynman diagrams for the elastic $2\rightarrow2$ scattering processes are shown in Fig.~\ref{collision_feynman_graphs}. Quark–gluon scattering receives contributions from three topologically distinct diagrams [Fig.~\ref{collision_feynman_graphs}(a)–Fig.~\ref{collision_feynman_graphs}(c)], while quark–quark and quark–antiquark scattering are described by a single diagram [Fig.~\ref{collision_feynman_graphs}(d)]~\cite{Svetitsky:1987gq}. In Fig.~\ref{collision_feynman_graphs}(a) and Fig.~\ref{collision_feynman_graphs}(d), the gluon propagator is replaced by the Gribov-modified propagator. In the Landau gauge, the Gribov gluon propagator is given by~\cite{Fukushima:2013xsa}
\begin{equation}
D_{\mu\nu}^{ab}(P) = \delta^{ab} \left( \delta_{\mu\nu} - \frac{P_\mu P_\nu}{P^2} \right) \frac{P^2}{P^4 + \gamma_G^4} \, . 
\label{propagator_GZ}
\end{equation}
(i) For the scattering $\mathrm{HQ}(P)+g(Q) \rightarrow$ $\mathrm{HQ}\left(P^{\prime}\right)+g\left(Q^{\prime}\right) $, the matrix elements reads as
\begin{subequations}
\begin{align}
\left|\mathcal{M}_{(a)}\right|^2 &=g_{\mathrm{HQ}} g_g\left[32 \pi^2 \alpha^2 \frac{\left(s-M_{\mathrm{HQ}}^2\right)\left(M_{\mathrm{HQ}}^2-u\right)t^{2}}{\left(t^{2}+\gamma_{\mathrm{G}}^4\right)^2}\right] \, , \\
\left|\mathcal{M}_{(b)}\right|^2 
&= g_{\mathrm{HQ}} g_g\left[\frac{64 \pi^2 \alpha^2}{9} \frac{\left(s-M_{\mathrm{HQ}}^2\right)\left(M_{\mathrm{HQ}}^2-u\right)+2 M_{\mathrm{HQ}}^2\left(s+M_{\mathrm{HQ}}^2\right)}{\left(s-M_{\mathrm{HQ}}^2\right)^2}\right] \, , \\
\left|\mathcal{M}_{(c)}\right|^2 
&= g_{\mathrm{HQ}} g_g\left[\frac{64 \pi^2 \alpha^2}{9} \frac{\left(s-M_{\mathrm{HQ}}^2\right)\left(M_{\mathrm{HQ}}^2-u\right)+2 M_{\mathrm{HQ}}^2\left(M_{\mathrm{HQ}}^2+u\right)}{\left(M_{\mathrm{HQ}}^2-u\right)^2}\right] \, , \\
\mathcal{M}_{(a)} \mathcal{M}_{(b)}^* 
&= \mathcal{M}_{(b)}^* \mathcal{M}_{(a)} = g_{\mathrm{HQ}}  g_g\left[8 \pi^2 \alpha^2 \frac{\left(s-M_{\mathrm{HQ}}^2\right)\left(M_{\mathrm{HQ}}^2-u\right)+M_{\mathrm{HQ}}^2(s-u)}{\left(\frac{t^{2}+\gamma_{\mathrm{G}}^4}{t} \right)\left(s-M_{\mathrm{HQ}}^2\right)}\right] \, , \\
\mathcal{M}_{(a)} \mathcal{M}_{(c)}^* 
&= \mathcal{M}_{(c)}^* \mathcal{M}_{(a)}= g_{\mathrm{HQ}} g_g\left[8 \pi^2 \alpha^2 \frac{\left(s-M_{\mathrm{HQ}}^2\right)\left(M_{\mathrm{HQ}}^2-u\right)-M_{\mathrm{HQ}}^2(s-u)}{\left(\frac{t^{2}+\gamma_{\mathrm{G}}^4}{t} \right)\left(M_{\mathrm{HQ}}^2-u\right)}\right] \, , \\
\mathcal{M}_{(b)} \mathcal{M}_{(c)}^* &=\mathcal{M}_{(b)}^* \mathcal{M}_{(c)}= g_{\mathrm{HQ}} g_g\left[\frac{8 \pi^2 \alpha^2}{9} \frac{M_{\mathrm{HQ}}^2\left(4 M_{\mathrm{HQ}}^2-t\right)}{\left(s-M_{\mathrm{HQ}}^2\right)\left(M_{\mathrm{HQ}}^2-u\right)}\right] \, , \\
\left|\mathcal{M}_{(i)}\right|^2 &= \left|\mathcal{M}_{(a)}\right|^2+\left|\mathcal{M}_{(b)}\right|^2+\left|\mathcal{M}_{(c)}\right|^2+2 \mathcal{R} e\left\{\mathcal{M}_{(a)} \mathcal{M}_{(b)}^*\right\}+2 \mathcal{R} e\left\{\mathcal{M}_{(b)} \mathcal{M}_{(c)}^*\right\} +2 \mathcal{R} e\left\{\mathcal{M}_{(a)} \mathcal{M}_{(c)}^*\right\} \, ,
\end{align}
\end{subequations}
and (ii) for the scattering $\mathrm{HQ}(P)+lq(Q)/l\bar{q}(Q) \rightarrow$ $\mathrm{HQ}\left(P^{\prime}\right)+lq\left(Q^{\prime}\right)/l\bar{q}\left(Q^{\prime}\right)$, one obtains
\begin{equation}
\begin{aligned}
\left|\mathcal{M}_{(d)}\right|^2= g_{\mathrm{HQ}} g_{lq/l \bar{q}}\left[\frac{64 \pi^2 \alpha^2}{9} \frac{\left\{\left(s-M_{\mathrm{HQ}}^2\right)^2+\left(M_{\mathrm{HQ}}^2-u\right)^2 + 2 M_{\mathrm{HQ}}^2 \frac{t^{2}+\gamma_{\mathrm{G}}^4}{t} \right\}t^{2}}{\left(t^{2}+\gamma_{\mathrm{G}}^4\right)^2}\right] \, .
\end{aligned}
\end{equation}
Here, $g_{\mathrm{HQ}} = N_{s}\times N_{c},\,  g_{g} = N_{s} \times (N_{c}^{2}-1)$ and $g_{lq/l \bar{q}} = N_{s}\times N_{c} \times N_{f}$ are the statistical degeneracy for the $\mathrm{HQ}$, gluon, and light quark particles respectively with $N_{s}=2, N_{f} =3$ and $N_{c} = 3$ have been considered. 
\section{PROJECTIONS OF ANISOTROPY VECTORS IN THE COM}\label{appendix_B}
The various projection mentioned in Eq.~\eqref{n_tilde_p_prime_avg.} are defined as follow as
\begin{subequations}
\begin{align}
\left({\bm{x}}_{c m} \cdot \bm{n}\right)&=\frac{\gamma_{c m}}{{p}_{c m}}\left[p \cos \Theta_{n}-E_p \frac{(p \cos \Theta_{n}+q \cos \chi \cos \Theta_{n}+q \sin \chi \cos \Phi \sin \Theta_{n})}{E_p+E_q}\right], \\
\left({\bm{y}}_{c m} \cdot \bm{n}\right)&=N^{-1}\left[\frac{p \cos \Theta_{n}+q \cos \chi \cos \Theta_{n}+q \sin \chi \cos \Phi \sin \Theta_{n}}{E_p+E_q}\right] \nn
&-\left(\bm{v}_{c m} \cdot {\bm{p}}_{c m}\right) \frac{\gamma_{c m}}{{p}_{c m}}\left[p \cos \Theta_{n}-E_p \frac{(p \cos \Theta_{n}+q \cos \chi \cos \Theta_{n}+q \sin \chi \cos \Phi \sin \Theta_{n})}{E_p+E_q}\right], \\
\left({\bm{z}}_{c m} \cdot \bm{n}\right)&=\gamma_{c m} N^{-1} \frac{1}{{p}_{c m}\left(E_p+E_q\right)} p q \sin \chi \sin \Phi \sin \Theta_{n}, \\
\left({\bm{x}}_{c m} \cdot \bm{p}\right)&=\frac{\gamma_{c m}}{{p}_{c m}}\left[p^2-E_p \frac{\left(p^2+p q \cos \chi\right)}{\left(E_p+E_q\right)}\right], \\
\left({\bm{y}}_{c m} \cdot \bm{p}\right)&=N^{-1}\left[\frac{p^2+p q \cos \chi}{\left(E_p+E_q\right)}-\left(\bm{v}_{c m} \cdot {\bm{p}}_{c m}\right) \frac{\gamma_{c m}}{{p}_{c m}^2}\left(p^2-\frac{E_p \left(p^2+p q \cos \chi\right)}{E_p+E_q}\right)\right].
\end{align}
\end{subequations}
\end{figure*}

\end{document}